\documentclass[aps,prd,preprint,fleqn,showkeys,showpacs,superscriptaddress,floatfix]{revtex4}
\usepackage[colorlinks=true]{hyperref}
\hypersetup{colorlinks=true, citecolor=blue, urlcolor=blue, linkcolor=blue}

\usepackage{graphicx}
\usepackage{tabulary}
\usepackage{multirow,tabularx}
\usepackage{amssymb}
\usepackage{amsmath}
\usepackage{bm}
\usepackage{enumitem}
\usepackage{etoolbox}

\usepackage{xcolor}
\makeatletter

\patchcmd\frontmatter@PACS@format{\addvspace{11\p@}}{}{}{}
\pretocmd\frontmatter@keys@format{\addvspace{11\p@}}{}{}
\patchcmd{\titleblock@produce}
  {\@pacs@produce\@pacs\@keywords@produce\@keywords}
  {\@keywords@produce\@keywords\@pacs@produce\@pacs}
  {}{}
\makeatother

\begin{document}
\title{Study of charm and beauty mass spectra, semileptonic decays of $B_{(s,c)}$ and ${B_c} \to J/\psi ({\eta _c}) + P(V)$ in a phenomenological potential model}

\author{S. Rahmani}
\email{s.rahmani120@gmail.com}
\affiliation{School of Physics and Electronics, Central South University, Changsha 410083, China}

\author{W. C. Luo}
\email{luo.wenchen@csu.edu.cn}
\affiliation{School of Physics and Electronics, Central South University, Changsha 410083, China}

\author{C. W. Xiao}
\email{xiaochw@csu.edu.cn}
\affiliation{School of Physics and Electronics, Central South University, Changsha 410083, China}

\begin{abstract} 
Using a non-relativistic potential model, we obtain the mass spectra, leptonic decay constants, and parameters of the Isgur-Wise function for the beauty and charm mesons. With the calculated quantities, we investigate purely leptonic decays of $B^+$, $B^{*}$, $B_c^+$, semileptonic decay modes ${B_{(s)}} \to {D_{(s)}}l\nu $  and  ${B_c} \to {\eta _c}\ell \bar \nu $ for three lepton channels $e,\mu ,\tau$, and obtain the corresponding branching fractions. $\bar B_{(s)}^{} \to D_{(s)}^*l\bar \nu $  transitions are also studied. Next, we apply the form factors for spin zero and spin one transitions of $B_c$ to calculate the nonleptonic branching ratios of ${B_c} \to J/\psi ({\eta _c}) + P(V)$, where $P$ and $V$ stand for the $D_q^{* - }$ vector meson and the $D_q$ pseudoscalar meson, respectively. Our results are found to be in agreement with those obtained in the experimental and theoretical results.
\end{abstract}

\maketitle
\thispagestyle{empty}

\section{Introduction }
Study of the weak decays of the $B$ mesons provides us with a good knowledge of the heavy quark dynamics and tests standard model predictions in this sector. The semileptonic decays of heavy hadrons are important subjects since they can be used to evaluate the CKM matrix element. In fact, the semileptonic heavy to heavy decays $B \to D\ell \nu $ in which they are connected with the $|{V_{cb}}|$, involve only one hadronic current that can be defined with the scalar functions known as form factors. These functions are parameterized by the hadronic matrix elements of the weak currents. Besides, the nonleptonic decays of the $B$ mesons are corresponding to the product of two single current matrix elements within the factorization approximation, and then the nonleptonic decay problem leads to the calculation of the meson form factors and the leptonic meson decay constants \cite{Neubert:1997uc}. The Isgur-Wise function (IWF) is useful to analyze the form factors of semileptonic and nonleptonic transitions of mesons and baryons including bottom and charmed. The weak decay modes of the $B$, $B_s$, and also $B_c$ meson, as a particle consisting of two heavy quarks, have been studied in various theoretical approaches. To study $B$ to $D$ meson decays, the IWF had been applied by \cite{Ligeti:1993hw}. Choi and Ji studied the exclusive nonleptonic two-body decays ${B_c} \to {D_{(s)}},{\eta _c}$  plus a pseudoscalar or vector meson using the QCD factorization approach \cite{Choi:2009ym}. Chen {\it{et al.}} analyzed the form factors of weak transitions $B_s^0 \to D_s^{(*) - }$  to obtain the decay properties of $B_s$ \cite{Chen:2011ut}. Na {\it{et al.}} found the ratio $R(D) = BR(B \to D\tau {\nu _\tau })/BR(B \to Dl{\nu _l}) = 0.300$  based on a lattice QCD calculation \cite{Na:2015kha}. Within the framework of the covariant conﬁned quark model, Dubnicka {\it{et al.}} calculated the branching fractions of ${B_c} \to J/\psi ({\eta _c}) + P(V)$ \cite{Dubnicka:2017job}. Zhou {\it{et al.}} applied the Bethe-Salpeter method to the semileptonic and nonleptonic decays of $B$, $B_s$, and $B_c$ mesons and obtained the branching ratios of the excited mesons \cite{Zhou:2020ijj}. Recently, the ATLAS Collaboration measured various branching fractions of the decays ${B_c} \to J/\psi D_s^ + $  and ${B_c} \to J/\psi D_s^{* + }$, where they reported the ratio: $BR({B_c} \to J/\psi D_s^{* + })/BR({B_c} \to J/\psi D_s^ + ) = 2.8_{ - 0.8}^{ + 1.2} \pm 0.3 $ \cite{ATLAS:2015jep}. We follow the covariant confined quark model for the helicity form factors of two-body nonleptonic decays of $B_c$ \cite{Dubnicka:2017job} and use the form factors which are related to our formalism, based on the parameterization near zero recoil point, to calculate the branching ratios. Since one of the important ingredients for the IWF is the wave function of the mesonic systems, we investigate the corresponding wave function, where the potential quark model is used to present a convenient wave function. 
\par
The leptonic decay constants of the beauty and charm mesons are important parameters in many processes, such as the ${B^0} - {\bar B^0}$ mixing, the nonleptonic decays, the determination of the CKM matrix element, and the examination of the lepton flavor universality. 
Generally, the purely bottom leptonic decays have simpler physics than the hadronic ones and they are related to the bound states of $B$ mesons, the CKM matrix elements, and the decay constants, which are closely connected with the quark-antiquark wave functions at the origin and can provide us the opportunity to study the microstructure and properties of mesons. The spectroscopy and leptonic decay constants of the $B$ and $D$ mesons are well calculated in different approaches. In 1976, Gershtein and Khlopov studied the leptonic decays of heavy pseudoscalar mesons with the quark model \cite{Gershtein:1976aq}. Yang obtained the wave functions and decay constants of $B$ and $D$ mesons using the relativistic potential model \cite{Yang:2011ie}. Mutuk calculated the mass spectra and decay constants of pseudoscalar and vector heavy-light mesons using the QCD sum rule and quark model \cite{Mutuk:2018lki}. Gutierrez-Guerrero {\it{et al.}} presented the mass spectra of mesons with one and  two heavy quarks in the charm and bottom sector based on a non-relativistic potential model by solving Schrödinger equation numerically \cite{Gutierrez-Guerrero:2021fuj}. Yang {\it{et al.}} studied the vector decay constants ${f_{{D^*}}},{f_{D_s^*}},{f_{{B^*}}}$, and ${f_{B_c^*}}$ in Ref. \cite{Yang:2021crs}.
\par
The paper is organized as follows. In section \ref{sec2}, we use the Killingbeck potential and the Gaussian wave function to obtain mass and leptonic decay constants. In Section \ref{sec3}, we study the IWF parameters, Semileptonic decay widths of $B$, $B_s$, and $B_c$ to charmed mesons. The nonleptonic decay modes of $B_c$ to charmed ones are considered in section \ref{sec4} and finally, we present a conclusion in section \ref{sec5}.
\section{Theoretical Framework}
\label{sec2}
\begin{figure}
\centering
\includegraphics[width=0.70\linewidth]{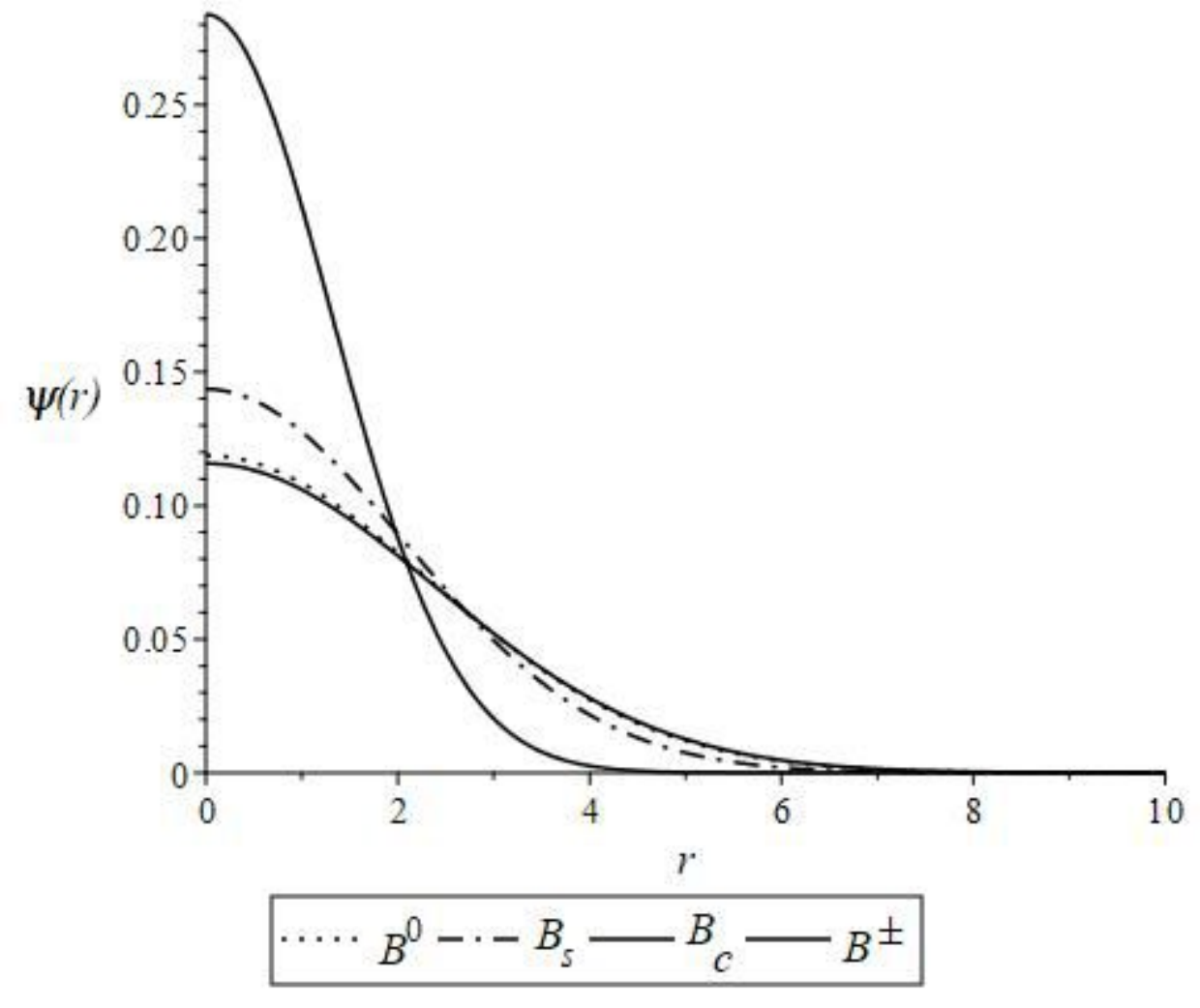}
\caption{Behavior of wave functions for the $B$ mesons.}
\label{fig:wave functions}
\end{figure}

In the non-relativistic quark model, the mesons are described as the bound states of quarks. In this framework, the interactions of constituent quarks are assumed in terms of potentials, which are taken phenomenologically inspiring by QCD. Since the kinetic energy of the quark constituents in a heavy-light meson system are small compared to their rest energy, the non-relativistic Hamiltonian is reasonable. The non-relativistic Hamiltonian of the mesons can be considered as
\begin{equation}
  H = \frac{{{p^2}}}{{2\mu }} + V(r),  
  \label{ham}
\end{equation}
where $p$ is the relative momentum of the quark-antiquark system, $\mu$  the reduced meson mass and the quark-antiquark potential given by 
\begin{equation}
   V(r) =  - \frac{4}{3}\frac{{{\alpha _s}}}{r} + br + a{r^2} + {V_0},
   \label{potential}
\end{equation}
which is the combination of the well-known Cornell potential plus harmonic, and often called the Killingbeck potential. In this potential, the Coulomb term ($- \frac{4}{3}\frac{{{\alpha _s}}}{r}$) is liable for the interaction at small distance originated from a Lorentz vector exchange and the terms ($br + a{r^2}$) take the responsibility of phenomenological confinement concept. The potential Eq. (\ref{potential}) is used for the properties of hadrons including mass spectra and decay widths \cite{Kumar:2013dsa,Hassanabadi:2014kka}. The potential parameters, $a$,   $b$, and $V_0$ can be determined by fitting to the experimental spectrum of mesons. $a$ is proportional to reduced mass and $\kappa $, via the relation $a = \frac{1}{2}\mu \kappa^2$ in which $\mu $ the reduced mass of the particle oscillates with the frequency $\kappa $. ${\alpha _s}$, the strong running coupling constant is determined as
\begin{equation}
    {\alpha _s} = \frac{{4\pi }}{{(11 - \frac{2}{3}{n_f})\ln (\frac{{{{(2\mu )}^2} + 1}}{{\Lambda _{QCD}^2}})}},
\end{equation}
with ${\Lambda _{QCD}} = 0.413$ GeV and the number of flavors $n_f = 3$. We use the Gaussian wave function in position space as \cite{Pang:2017dlw}
\begin{equation}
{\psi _{n,l}}(g,r) = N{g^{\frac{3}{2} + l}}{r^l}{e^{ - {g^2}{r^2}/2}}L_{n - 1}^{l + 1/2}({g^2}{r^2}),
\label{eq:wave}
\end{equation}
where $g$ is the variational parameter, $N$ normalization constant and $L_{n - 1}^{l + 1/2}({g^2}{r^2})$ the Laguerre polynomial. $g$ can be obtained by minimizing the trial energy \cite{Xiao:2020gry}. The Harmonic oscillator basis for the wave functions has been applied to hadronic systems in different literature. Roberts and Pervin considered the harmonic oscillator wave function for a potential consisting of linear and Coulomb components \cite{Roberts:2007ni}. Kumar and Chand solved the radial Schrödinger equation for the Killingbeck potential by choosing the wave function as a Gaussian-type function, $\exp ( - \alpha {r^2} - \beta r)$ \cite{Kumar:2013dsa}. Note that our trial wave function in Eq. (\ref{eq:wave}) is consistent with them. It should be mentioned that the Laguerre polynomials are proportional to the exponential function $\exp ( - \beta r)$. Bhaghyesh {\it{et al.}} used the harmonic oscillator wave function via a variational approach to obtain the mass spectrum of charmonium and bottomonium \cite{Bhaghyesh:2011zza}, where they considered the Hulthen potential plus a linear confining term. In Ref. \cite{VijayaKumar:2004hn}, three-dimensional harmonic oscillator wave functions were taken as the trial wave functions where the harmonic along with one-gluon-exchange potential 
were used to obtain the meson masses \cite{VijayaKumar:2004hn}. Pang chose simple harmonic oscillator basis wave functions considered in the quark model with the screening potential \cite{Pang:2019ttv}.
The behaviors of wave functions for different $B$ mesons are shown in Fig. \ref{fig:wave functions}. In this figure, $B$, $B_s$, $B_c$, and $B^0$ are referred to as the ground states. The radial wave functions of the mentioned bottom mesons in this figure have been plotted along the quark-antiquark distance $r$ ($GeV^{-1}$) for the case of $n=1$ mode, which are defined by Eq. (\ref{eq:wave}) and where they drop to zero since the quark and antiquark bound in the mesons. Hence, one can expect that the mesonic wave functions diminish at a typically large distance.

We can calculate the masses of charm and bottom mesons using
\begin{equation}
    {M_{P/V}} = {m_1} + {m_2} + E_{1,0} + \left\langle {{V_{sd}}} \right\rangle,
\end{equation}
where ${m_1},{m_2}$ and $E_{1,0}$ are the quark masses and the energy of the mesons respectively. Using the Hamiltonian of Eq. (\ref{ham}), the wave function of Eq. (\ref{eq:wave}) as well as the expectation value of the Hamiltonian, ${E_{1,0}} = \left\langle {{\psi _{1,0}}(g,r)|H|{\psi _{1,0}}(g,r)} \right\rangle $, we can get the energy of the mesons \cite{Xiao:2020gry}. The quark masses are taken as ${m_u} = 0.338, {m_d} = 0.350, {m_s} = 0.469, {m_b} = 4.98, {m_c} = 1.5$, which are all in unit GeV and determined by fitting to the spectrum of mesons. $V_{sd}$ stands for the spin-dependent interaction, given by
\begin{equation}
    {V_{sd}} = \frac{2}{{3{m_1}{m_2}}}({\vec s_1}.{\vec s_2})(4\pi {\alpha _s}\delta (r) + 6A),
\end{equation}
where we take $A = 0.014GeV$ \cite{Lengyel:2000dk}, and use the approximation of 
\begin{equation}
    \delta (r) = \frac{{{{\omega '}^2}}}{{\pi r}}{e^{ - 2\omega 'r}},
\end{equation}
with ${\omega '^2} = \frac{{2m_1^2m_2^2}}{{m_1^2 + m_2^2}}$ \cite{Radford:2009bs} and the notation $\left\langle {{\vec s}_1}.{{\vec s}_2} \right\rangle  = \frac{1}{2}[S(S + 1) - {s_1}({s_1} + 1) - {s_2}({s_2} + 1)]$. By fitting to the experimental masses of $B$ and $D$ mesons, we take the potential parameters as $\kappa = 0.24 GeV$ and $b = 0.10 GeV^2$, where one should keep in mind that $a = \frac{1}{2}\mu {\kappa ^2}$. The calculated masses for $B$ mesons are shown in Table \ref{tab:massb}. The first column shows the considered ground and excited states of $B$ mesons. The second column is the obtained variational parameter. The next columns stand for our masses of mesons and the experimental ones.  We also tabulate the masses of charm mesons in Table \ref{tab:massc}. The root of mean square deviations can be obtained by
\begin{equation}
    \sigma  = \sqrt {\frac{1}{{N'}}\sum\limits_{n = 1}^{N'} {{{\left[ {\frac{{{M_{theor}} - {M_{\exp }}}}{{{M_{\exp }}}}} \right]}^2}} }. 
\end{equation}
We obtain 0.002 for the deviation of masses of pseudoscalar $B$ mesons, and 0.009 for the deviation of pseudoscalar charm mesons. The mass differences of  $B$ and $B^*$, $B_{s}$ and $B_s^*$, $B_c^ + $ and $B_{s}$ have been reported as ${M_{{B^*}}} - {M_B} = 45.21 \pm 0.21{\text{MeV}}$, ${M_{B_s^*}} - {M_{{B_s}}} = 48.5_{ - 1.5}^{ + 1.8}$ MeV and ${M_{B_c^ + }} - {M_{B_s^0}} = 907.8 \pm 0.5$ MeV \cite{PDG:2022}. In our approach, we obtain ${M_{{B^*}}} - {M_B}$ = 57.02 MeV, ${M_{B_s^*}} - {M_{{B_s}}}$= 51.25 MeV and ${M_{B_c^ + }} - {M_{B_s^0}} = $ 926.50 MeV, which are in compatible with these results of \cite{PDG:2022}. The differences of our calculated masses with experimental reports for the bottom sector are -8.97 MeV, -22.64 MeV, -3.68 MeV, -4.22 MeV, 2.68 MeV, and -19.87 MeV for the states ${B^ \pm }(u\bar b)$, $B_s^0(s\bar b)$, $B_c^ + (c\bar b)$, ${B^0}(d\bar b)$, ${B^*}(u\bar b)$ and $B_s^{0*}(s\bar b)$, respectively. For the case of charm sector, the associated differences of ${D^0}(c\bar u)$, ${D^ \pm }(c\bar d)$, $D_s^ + (c\bar s)$, ${\eta _c}(1S)$, ${D^*}$, $D_s^{* \pm }$ and $J/\psi (1S)$ are 6.81 MeV, 8.97 MeV, -7.98 MeV, -51.79 MeV, 45.21 MeV, 4.47 MeV, and 76.13 MeV, respectively. We tabulate the errors of our results in the final columns of Tables \ref{tab:massb} and \ref{tab:massc}.

\begin{center}
\begin{table}
\caption{Masses of bottom mesons ($V_0=-0.35$ GeV).}
\label{tab:massb}
\begin{tabular}{|p{2.5cm}|p{2.5cm}|p{2.5cm}|p{2.5cm}|p{2.5cm}|}
\hline    
\textbf{Meson} &  \textbf{\textit{g}}  & \textbf{Our mass (GeV)} &   \textbf{Exp. mass \cite{PDG:2022}}  & \textbf{Error ($\%$)}\\ \hline  
${B^ \pm }(u\bar b)$ & {0.421}	 & {5.270} & {5.279} & {0.17} \\ \hline
$B_s^0(s\bar b)$ & {0.486} & {5.344} & 5.367 & {0.42}\\ \hline
$B_c^ + (c\bar b)$ & {0.766} & {6.271} & 6.274 & {0.06} \\ \hline
${B^0}(d\bar b)$ & {0.428} & {5.275} & 5.280 & {0.08} \\ \hline
${B^*}(u\bar b)$ & {0.421} & {5.327} & 5.325 & {0.05}\\ \hline
$B_s^{0*}(s\bar b)$ & 0.486 & {5.396} & 5.415 & {0.37}\\ \hline

\end{tabular} 
\end{table}
\end{center}

\begin{center}
\begin{table}
\caption{Masses of different states of $D$, $D_s$ and charmonium mesons {($V_0=-0.21$ GeV)}.}
\label{tab:massc}
\begin{tabular}{|p{2.5cm}|p{2.5cm}|p{2.5cm}|p{2.5cm}|p{2.5cm}|}
\hline    
\textbf{Meson} &  \textbf{\textit{g}}  & \textbf{Our mass (GeV)} &   \textbf{Exp. mass \cite{PDG:2022}} & \textbf{{Error ($\%$)} } \\ \hline  
${D^0}(c\bar u)$ & {0.394} & {1.872} & {1.865} & {0.37}\\ \hline
${D^ \pm }(c\bar d)$ & {0.399} & {1.879} &{ 1.870} & {0.48} \\ \hline
$D_s^ + (c\bar s)$ &{0.446} & {1.960} & {1.968} & {0.41}\\ \hline
${\eta _c}(1S)$ & {0.628} & {2.932} & { 2.984} & {1.74} \\ \hline
${D^*}$ & {0.399} & {2.055} & {2.010} & {2.25} \\ \hline
$D_s^{* \pm }$ & {0.446} & {2.117} & {2.112} & {0.21} \\ \hline
$J/\psi (1S)$ & {0.628} & {3.021} & {3.096} & {2.46} \\ \hline
\end{tabular} 
\end{table}
\end{center}

In next step, we calculate the leptonic decay constants based on the non-relativistic formula, written as
\begin{equation}
 f_{p/v}^2 = \frac{{12|\psi (0){|^2}}}{{{M_{p/v}}}},
 \label{eq:decay constant without QCD}
\end{equation}
where  $|\psi (0)|$ is the wave function at the origin which can be obtained through the standard relation $|\psi (0){|^2} = \frac{\mu }{{2\pi }}\left\langle {\frac{{dV(r)}}{{dr}}} \right\rangle$, and $\left\langle {\frac{{dV(r)}}{{dr}}} \right\rangle $ is the expectation value of ${\frac{{dV(r)}}{{dr}}}$ and taken by the wave function, Eq. (\ref{eq:wave}). We obtain the square of the wave function at the origin and decay constant as shown in the second and third columns of Table \ref{tab:leptonic}. We also include the QCD correction factor $\bar C({\alpha _s})$, which is given by \cite{Braaten:1995ej}
\begin{equation}
    {\bar C^2}({\alpha _s}) = 1 - \frac{{{\alpha _s}}}{\pi }[{\Delta _{P/V}} - \frac{{{m_Q} - {m_{\bar q}}}}{{{m_Q} + {m_{\bar q}}}}\ln \frac{{{m_Q}}}{{{m_{\bar q}}}}],
\end{equation}
with ${\Delta _P} = 2$  and ${\Delta _V} = \frac{8}{3}$ by multiplying in the leptonic decay constant  
\begin{equation}
    \bar f_{p/v}^2 = \frac{{12|\psi (0){|^2}}}{{{M_{p/v}}}}{\bar C^2}({\alpha _s}).
    \label{eq:decay constant with QCD}
\end{equation}
The fourth column of Table \ref{tab:leptonic} shows our results for the leptonic decay constants of pseudoscalar and vector mesons with QCD correction. In the final column, the results of Refs. \cite{Belle:2006but, Sun:2019xyw, Pathak:2011km, Eichten:2019gig, Ebert:2002pp, Colquhoun:2015oha, Wang:2005qx, Albertus:2005vd, Ebert:2006hj, Dubnicka:2017job, Bhaghyesh:2011zza} are presented for comparisons. The ratios of decay constants are $\frac{{{f_{{B^*}}}}}{{{f_B}}}{\text{ = 0}}{\text{.927}}$ and $\frac{{{f_{B_s^*}}}}{{{f_{{B_s}}}}} = 0.{\text{929}}$ using the values of fourth column for the bottom section. Colquhoun {\it{et al.}} reported these quantities as $\frac{{{f_{B_{}^*}}}}{{{f_B}}} = 0.941$ and $\frac{{{f_{B_s^*}}}}{{{f_{{B_s}}}}} = 0.953$ \cite{Colquhoun:2015oha}. For the ratios of charm sector, we evaluate  $\frac{{{f_{D_s^*}}}}{{{f_{{D^*}}}}} = {\text{1.}}{\text{139}}$, which is in comparable with $\frac{{{f_{D_s^*}}}}{{{f_{{D^*}}}}} = 1.12$ obtained by Chang {\it{et al.}} \cite{Chang:2018aut}.

\begin{center}
\begin{table}
\caption{Leptonic decay constants.}
\label{tab:leptonic}
\begin{tabular}{|p{2.5cm}|p{2.5cm}|p{2.5cm}|p{3.1cm}|p{2.5cm}|}
\hline    
\textbf{Meson} &  \textbf{${|\psi (0){|^2}({GeV^3)}}$}  & \textbf{${f_{p/v}}$({MeV})} &  \textbf{${{\bar f}_{p/v}}$ (MeV)} & \textbf{Others (MeV)}  \\ \hline  
${B^ \pm }(u\bar b)$ & {0.023} & {230} & {239} & $229_{ - 31 - 37}^{ + 39 + 34}$ \cite{Belle:2006but}, {210 \cite{Sun:2019xyw}}\\ \hline
$B_s^0(s\bar b)$ & {0.037} & {287} & {285} & 265 \cite{Pathak:2011km} \\ \hline
$B_c^ + (c\bar b)$ & {0.147} & {531} & {484} & 498 \cite{Eichten:2019gig}, 489 \cite{Dubnicka:2017job}, 433 \cite{Ebert:2002pp}, 434 \cite{Colquhoun:2015oha}\\ \hline
${B^0}(d\bar b)$ & 0.025 & 236 & 244 & 246 \cite{Pathak:2011km}, 213 \cite{Pathak:2011km} \\ \hline
${B^*}(u\bar b)$ & 0.023 & 229 & 221 & $238 \pm 18$ \cite{Wang:2005qx}, 175 \cite{Colquhoun:2015oha}, 223 \cite{Sun:2019xyw} \\ \hline
$B_s^{0*}(s\bar b)$ & 0.037 & 285 & 265 & $236_{ - 11}^{ + 14}$ \cite{Albertus:2005vd}, 242 \cite{Sun:2019xyw} \\ \hline
${D^0}(c\bar u)$ & 0.019 & 348 & 305 & -  \\ \hline
${D^ \pm }(c\bar d)$ & 0.020 & {355} & {310} & 376 \cite{Ebert:2006hj}, $243_{ - 17}^{ + 21}$ \cite{Albertus:2005vd} \\ \hline
$D_s^ + (c\bar s)$ & {0.028} & {414} & {350} & 436 \cite{Ebert:2006hj}, $341_{ - 5}^{ + 7}$ \cite{Albertus:2005vd}\\ \hline
${\eta _c}(1S)$ & {0.081} & {575} & {480} & 628 \cite{Dubnicka:2017job}, {471 \cite{Bhaghyesh:2011zza}} \\ \hline
${D^*}$ & {0.020} & {339} & {267} & $223_{ - 19}^{ + 23}$ \cite{Albertus:2005vd}\\ \hline
$D_s^{* \pm }$ & {0.028} & {398} & {303} & $326_{ - 17}^{ + 21}$ \cite{Albertus:2005vd}\\ \hline
$J/\psi (1S)$ & {0.081} & {566} & {438} & 415 \cite{Dubnicka:2017job}, {462 \cite{Bhaghyesh:2011zza}} \\ \hline
\end{tabular} 
\end{table}
\end{center}

\par
We calculate the branching ratios of purely leptonic decays of $B^+$, ${B^{* \pm }}$ and $B_c^+$ using the obtained masses and leptonic decay constants from Tables \ref{tab:massb} and \ref{tab:leptonic}. These branching ratios can be written as
\begin{equation}
    BR(B_q^ \pm  \to {l^ \pm }{\nu _l}) = \frac{{G_F^2m_l^2{M_{{B_q}}}}}{{8\pi }}{\left( {1 - \frac{{m_l^2}}{{M_{{B_q}}^2}}} \right)^2} \times {f_{{p/v}}^2}|{V_{qb}}{|^2}{\tau _{{B_q}}},q = u,c,
\end{equation}
\begin{equation}
    BR(B_{}^{* \pm } \to {l^ \pm }{\nu _l}) = \frac{{G_F^2M_{{B^{* \pm }}}^3}}{{12\pi }}\left( {1 - \frac{3}{2}\frac{{m_l^2}}{{M_{{B^{* \pm }}}^2}} + \frac{1}{2}\frac{{m_l^6}}{{M_{{B^{* \pm }}}^6}}} \right) \times {f_{p/v}^2}|{V_{ub}}{|^2}{\tau _{{B^{* \pm }}}},
\end{equation}
where the decay amplitudes are dominated by the tree-level diagrams, $G_F$ is the Fermi coupling constant, $m_l$ the lepton mass, $M_{B}$ and $M_B^{*\pm}$ the masses of pseudoscalar and vector $B$ mesons, $V_{qb}$ the CKM matrix elements, $\tau_B$ and $\tau _{B^{* \pm }}$ the lifetimes of pseudoscalar and vector $B$ mesons. Since the total decay widths of vector $B$ meson are dominated by the electromagnetic processes and the other decay
modes are too rare for experimental measurements, we take the value of total decay width of $B^*$ as ${\Gamma _{{B^{* + }}}} \approx \Gamma ({B^*} \to B\gamma) \approx 0.468$ keV \cite{Sun:2019xyw}. We show the obtained values for leptonic channels of the $B$ mesons in Table \ref{tab:br bottom leptonic}. The second and third columns are the results  using the decay constants as Eqs. (\ref{eq:decay constant without QCD}) and (\ref{eq:decay constant with QCD}), respectively, which are consistent with the results obtained in Ref. \cite{Sun:2019xyw}.

\begin{center}
\begin{table}
\caption{Purely leptonic branching ratios of $B^+$, ${B^{* \pm }}$ and $B_c^+$.}
\label{tab:br bottom leptonic}
\begin{tabular}{|p{2.5cm}|p{3cm}|p{3cm}|p{3cm}|}
\hline    
\textbf{Channel} &  \textbf{Ours (with ${f_{p/v}}$)}  & 
\textbf{Ours (with ${{\bar f}_{p/v}}$)} &  \textbf{Ref.} \cite{Sun:2019xyw}  \\ \hline  
$B^ +  \to {e^ + }\nu _e$ & {1.44 $\times 10^{ - 11}$} & {$1.54 \times 10^{ - 11}$} & {$1.27 \times 10^{-11}$}  \\ \hline
{${B^ + } \to {\mu ^ + }{\nu _\mu }$} & {${\text{6}}{\text{.13}} \times {10^{ - 7}}$} & {${\text{6}}{\text{.58}} \times {10^{ - 7}}$} & {$5.4 \times {10^{ - 7}}$}  \\ \hline
{${B^ + } \to {\tau ^ + }{\nu _\tau }$} & {${\text{1}}{\text{.36}} \times {10^{ - 4}}$} & {${\text{1}}{\text{.46}} \times {10^{ - 4}}$} & {$1.21 \times {10^{ - 4}}$} \\ \hline
{$B_c^ +  \to {e^ + }{\nu _e}$} & {${\text{3}}{\text{.22}} \times {10^{ - 9}}$} & {${\text{2}}{\text{.68}} \times {10^{ - 9}}$} & {$2.24 \times {10^{ - 9}}$} \\ \hline
{$B_c^ +  \to {\mu ^ + }{\nu _\mu }$} & {${\text{13}}{\text{.76}} \times {10^{ - 5}}$} & {${\text{11}}{\text{.47}} \times {10^{ - 5}}$} & {$9.6 \times {10^{ - 5}}$} \\ 
\hline
{$B_c^ +  \to {\tau ^ + }{\nu _\tau }$} & {${\text{3}}{\text{.29}} \times {10^{ - 2}}$}  & {${\text{2}}{\text{.74}} \times {10^{ - 2}}$} & {$2.29 \times {10^{ - 2}}$} \\
\hline
{${B^{* + }} \to {e^ + }{\nu _e}$} & {${\text{8}}{\text{.93}} \times {10^{ - 10}}$} & {${\text{8}}{\text{.33}} \times {10^{ - 10}}$}  & {${\text{9}}{\text{.0}} \times {10^{ - 10}}$} \\
\hline
{${B^{* + }} \to {\mu ^ + }{\nu _\mu }$} & {${\text{8}}{\text{.92}} \times {10^{ - 10}}$} & {${\text{8}}{\text{.32}} \times {10^{ - 10}}$} & {${\text{9}}{\text{.0}} \times {10^{ - 10}}$} \\
\hline
{${B^{* + }} \to {\tau ^ + }{\nu _\tau }$} & {${\text{7}}{\text{.44}} \times {10^{ - 10}}$} & {${\text{6}}{\text{.94}} \times {10^{ - 10}}$} & {${\text{7}}{\text{.5}} \times {10^{ - 10}}$} \\
\hline
\end{tabular} 
\end{table}
\end{center}

\section{Isgur-Wise function parameters, Semileptonic decay widths of $B$, $B_s $ and $B_c$  to charmed mesons }
\label{sec3}
\begin{figure}
\centering
\includegraphics[width=0.70\linewidth]{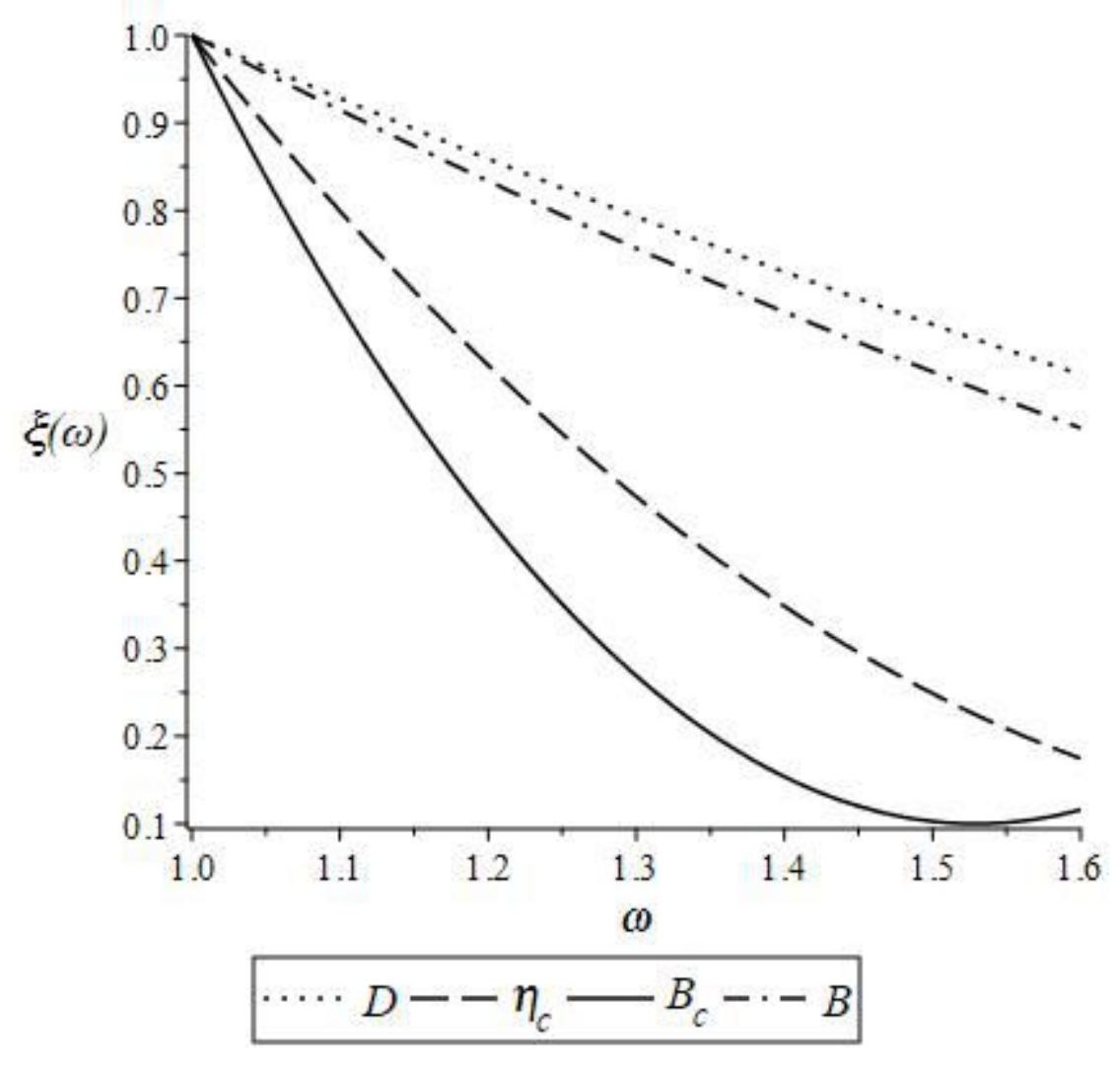}
\caption{ IWFs for charm and beauty mesons.}
\label{fig:IWF}
\end{figure}

Having the wave functions of mesons, Eq. (\ref{eq:wave}), we can evaluate the IWF and its parameters which are applied to the semileptonic and nonleptonic decays of heavy-heavy and heavy-light mesons. In fact, the form factors of semileptonic decays in heavy quark limit can be defined by the IWF. This form factor can be written in terms of the dot product of four velocities of initial and final mesons, $\omega $, as following
\begin{equation}
    \xi (\omega ) = 1 - {\rho ^2}(\omega  - 1) + C{(\omega  - 1)^2} ,
    \label{eq:IWF}
\end{equation}
where the slope and curvature of this function are given by \cite{Hassanabadi:2014isa,Rahmani:2017vbg}
\begin{equation}
\begin{gathered}
  {\rho ^2} = 4\pi {\mu ^2}\int\limits_0^\infty  {{r^4}} {({\psi _{1,0}}(g,r))^2}dr, \hfill \\
  C = \frac{2}{3}\pi {\mu ^4}\int\limits_0^\infty  {{r^6}} {({\psi _{1,0}}(g,r))^2}dr. \hfill \\ 
\end{gathered} 
\end{equation}
These parameters are originated from the momentum transfer ${p^2} = 2{\mu ^2}(\omega  - 1)$. In Table \ref{tab:IWF}, we show our results for the slopes and curvatures. We use these parameters for the bottom mesons in the next step. Since we are dealing with pseudoscalar decays, we show our parameters of the IWF for these cases of $0^-$ states. For the $D_s$ meson, the slope of the IWF has been reported in QCD lattice approach 1.19 \cite{Atoui:2013zza}. Our result of 0.96 is consistent with them \cite{Atoui:2013zza}. The slope parameter has been reported in a lattice QCD calculation as ${\rho ^2} = 1.119$ \cite{Na:2015kha}. We compare our results of slopes and curvatures for the beauty and charm mesons with Refs. \cite{PDG:2022, Faller:2008tr, Hassanabadi:2014isa, LHCb:2020hpv} in Table \ref{tab:IWF}. The behavior of IWFs are plotted in Fig. \ref{fig:IWF} for $B$, $D$, $B_c$ and $\eta_c$. As one can see from Fig. \ref{fig:IWF}, $B_c$ drops faster than other mesons due to its larger values of $\rho ^2$  and $C$. According to Eq. (\ref{eq:IWF}) and the calculated parameters of Table. \ref{tab:IWF}, we proceed with the semileptonic decays of bottom mesons.

\begin{center}
\begin{table}[ht]
\caption{Slopes and curvatures for different bottom and charm pseudoscalar mesons.}
\label{tab:IWF}
\begin{tabularx}{\textwidth}{|X|X|X|X|} 
\hline    

 \textbf{Meson}  & \textbf{${\rho ^2}$ (this work)}  & \textbf{$C$ (this work)} & \textbf{(Others)} \\ \hline  
   
${B^ \pm }(u\bar b)$  &0.85 & {0.20}  & ${\rho ^2} = 0.81 \pm 0.22$  \cite{Faller:2008tr}, ${\rho ^2} = 0.74,C = {\text{0}}{\text{.13}}$ \cite{Hassanabadi:2014isa} \\ \hline
{${B^0}(d\bar b)$} & {$0.87$} & {$0.21$} & - \\  \hline

$B_s^0(s\bar b)$  & {1.17} & {0.38}  & ${\rho ^2} = {\text{1}}{\text{.36}},C = 0.46$ \cite{Hassanabadi:2014isa}, ${\rho ^2} = 1.16 \pm 0.05 \pm 0.07$ \cite{LHCb:2020hpv}, ${\rho ^2} = 1.17 \pm 0.08$ \cite{PDG:2022}  \\ \hline
$B_c^ + (c\bar b)$  & {3.40} & {3.21} &  - \\ \hline
${D^0}(c\bar u)$ & {0.74} & {0.15} & ${\rho ^2} = 0.62,C = {\text{0}}{\text{.09}}$ \cite{Hassanabadi:2014isa} \\ \hline
{${D^ \pm }(c\bar d)$} & {0.76} & {0.16} & - \\ \hline
$D_s^ + (c\bar s)$ & {0.96} & {0.26} & ${\rho ^2} = {\text{1}}{\text{.06}},C = 0.28$ \cite{Hassanabadi:2014isa} \\ \hline
${\eta _c}(1S)$ & {2.14} & {1.27} & - \\ \hline
\end{tabularx} 
\end{table}
\end{center}

\par
\begin{figure}
\centering
\includegraphics[width=0.48\linewidth]{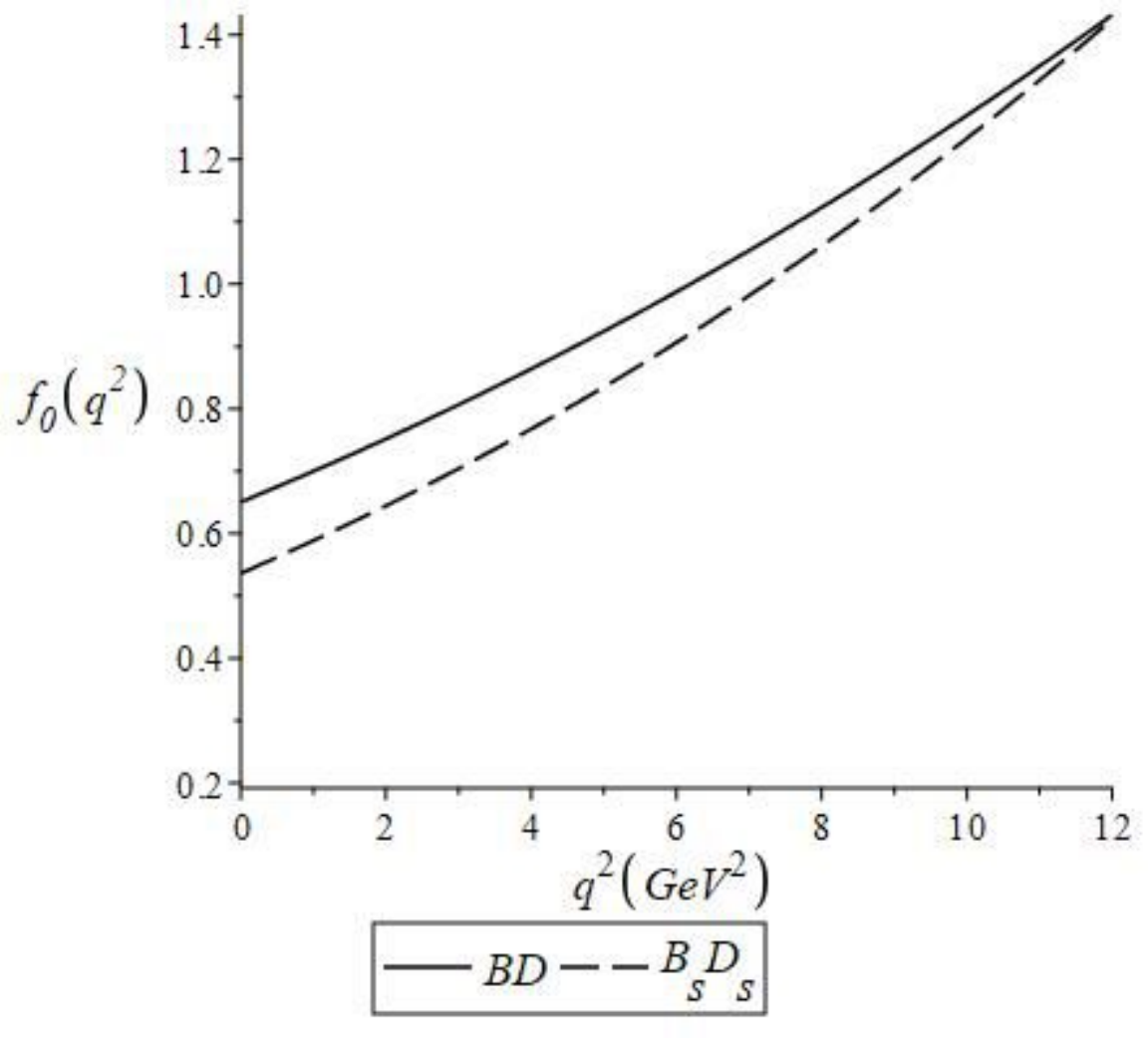}
\caption{The transfer momentum dependence of the form factor ${f_0}({q^2})$ for two semileptonic decays $B$ to $D$ and $B_s$ to $D_s$.}
\label{fig:f_0 BD}
\end{figure}

In terms of two transition form factors $f_{BD}^ + ({q^2})$ and $f_{BD}^0({q^2})$, which are associated to the decay $B \to Dl\nu $, the differential semileptonic decay width of $B \to Dl\nu $ can be obtained by  \cite{Wang:2017jow}
\begin{equation}
\label{eq:13semi}
    \begin{gathered}
  \frac{{d\Gamma (B \to Dl\nu )}}{{d{q^2}}} = \frac{{\eta _{EW}^2G_F^2|{V_{cb}}{|^2}}}{{24{\pi ^3}M_B^2}}{\left( {1 - \frac{{m_l^2}}{{{q^2}}}} \right)^2}|{{\vec p}_D}|[\left( {1 + \frac{{m_l^2}}{{2{q^2}}}} \right)M_B^2|{{\vec p}_D}{|^2}|f_{BD}^ + ({q^2}){|^2} \hfill \\
   + \frac{{3m_l^2}}{{8{q^2}}}{(M_B^2 - M_D^2)^2}|f_{BD}^0({q^2}){|^2}], \hfill \\ 
\end{gathered} 
\end{equation}
where the three momentum of $D$ meson is given by
\begin{equation}
    |{\vec p_D}| = \sqrt {\lambda (M_B^2,M_D^2,{q^2})} /(2{M_B}),
\end{equation} 
with $\lambda (M_B^2,M_D^2,{q^2})$ the K\"allen triangle function. The masses of mesons, $M_B$ and $M_D$ are taken from previous section and ${\eta _{EW}}{\text{ = 1}}{\text{.0066}}$. $G_F$ is Fermi coupling constant. We integrate Eq. (\ref{eq:13semi}) over  $q^2$  within the range of $m_l^2 \leqslant {q^2} \leqslant {({M_B} - {M_D})^2}$. We can write the form factors for the decay $B \to Dl\nu $ by employing Heavy Quark Effective Theory (HQET) \cite{Isgur:1990yhj, Xiao:2014ana, Choi:2021mni}
\begin{equation}
\label{eq:form factors}
    \begin{gathered}
{
  {f_ \pm (q^2) } = \xi (\omega )\frac{{{M_{{B}}} \pm {M_D}}}{{2\sqrt {{M_{{B}}}{M_D}} }},} \hfill \\
  {
  {f_0}({q^2}) = {f_ + }({q^2}) + \frac{{{q^2}}}{{M_B^2 - M_D^2}}{f_ - }({q^2}),} \hfill \\ 
\end{gathered} 
\end{equation}
where $\xi (\omega )$ is the IWF. In Eq. (\ref{eq:13semi}), we take the form factors  $f_{BD}^ + ({q^2})$ and $f_{BD}^0({q^2})$ as Eqs. (\ref{eq:form factors}), corresponding to ${f_ + (q^2) }$ and ${f_0}({q^2})$, respectively. We use the kinetic variable as $\omega  = \frac{{M_B^2 + M_D^2 - {q^2}}}{{2{M_B}{M_D}}}$  to obtain form factors. The input lepton masses are taken from \cite{PDG:2022}, ${m_\tau } = 1.776,{m_e} = 0.510 \times {10^{ - 3}},{m_\mu } = 0.105$ GeV, and the CKM matrix element is taken as $|{V_{cb}}| = 40.8 \times 10^{-3}$. Analogous relations hold for the semileptonic transitions of $B_s$ to $D_s$. In Fig. \ref{fig:f_0 BD}, we show the behavior of form factors for two semileptonic decay widths of $B$ and $B_s$. At $q^2$ =0, two form factors ${f_0}({q^2})$ and ${f_+}({q^2})$  are placed at 0.650 and 0.536 for the decays $B \to Dl\nu $ and $B_s \to {D_s}l\nu $, respectively. ${f_ + }(0) = 0.664$ for the semileptonic decay $B \to Dl\nu $ was reported in Ref. \cite{Na:2015kha}.
\par
\begin{figure}
\centering
\includegraphics[width=0.48\linewidth]{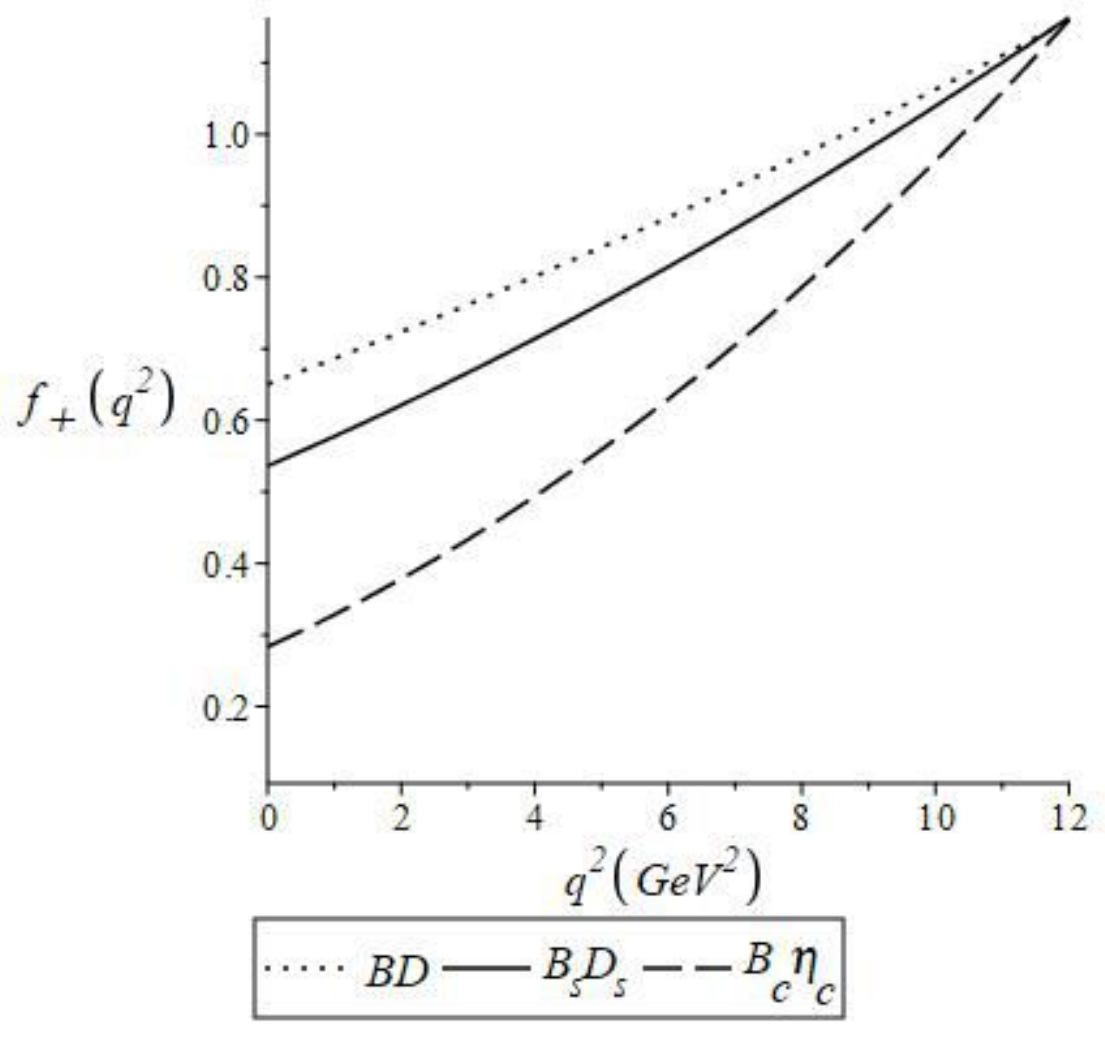}
\caption{ The transfer momentum dependence of the form factor ${f_ + }({q^2})$  for three bottom semileptonic decays $B$ to $D$, $B_s$ to $D_s$ and $B_c$ to $\eta_c$.}
\label{fig:f_plus BD}
\end{figure}

For the differential semileptonic decay width of $B_c$ to a charmonium pseudoscalar state $\eta_c$, ${B_c} \to {\eta _c}\ell \bar \nu $, we can write \cite{Rahmani:2017vbg}
\begin{equation}
\label{eq:16}
    \frac{{d\Gamma }}{{d{q^2}}}({B_c} \to {\eta _c}\ell \bar \nu ) = \frac{{G_F^2|{V_{cb}}{|^2}}}{{24{\pi ^3}}}{(\frac{{{{(M_{{B_c}}^2 + M_{{\eta _c}}^2 - {q^2})}^2}}}{{4M_{{B_c}}^2}} - M_{{\eta _c}}^2)^{\frac{3}{2}}}|{f_ + }({q^2}){|^2}.
\end{equation} 
Fig. \ref{fig:f_plus BD} depicts the transfer momentum dependence of form factors ${f_ + }({q^2})$ for the three mentioned semileptonic decays. One can see from Figs. \ref{fig:f_0 BD} and \ref{fig:f_plus BD}, two transition form factors, $f_+(q^2)$ and $f_0(q^2)$ have the same value $0.650$ at $q^2 = 0$ $GeV^2$ for the case of $B \to Dl\nu $ decay. For the decay $B_s \to D_sl\nu $, we have $f_+(0)$ = $f_0(0)$ = 0.536. At $q^2 = 0$ $GeV^2$, it is found that $f_+(q^2)=0.284$ in the semileptonic decay of $B_c$ to $\eta_c$. As we can see in Figs. \ref{fig:f_0 BD} and \ref{fig:f_plus BD}, the quantities $f_+(q^2)$ and $f_0(q^2)$ increase along with the transfer momentum. Our results for $f_+$ and $f_0$ are in agreement with those obtained in \cite{Choi:2021mni}, and also compatible with the lattice QCD calculations by the HPQCD collaboration \cite{Na:2015kha}, where the quantities $f_+$ and $f_0$ were plotted versus transfer momentum in the physical kinematic range of 0 to 12 $GeV^2$. Fig. \ref{fig:f_0 BD} grows faster with $q^2$ compared to the results obtained from the lattice calculations \cite{Na:2015kha}. Note that $f_+(q^2)$ is the vector form factor while $f_0(q^2)$ is the scalar form factor, which are obtained by Eq. (\ref{eq:form factors}), and one would expect that they are equal at the zero momentum transfer.

\begin{figure}
\centering
\includegraphics[width=0.60\linewidth]{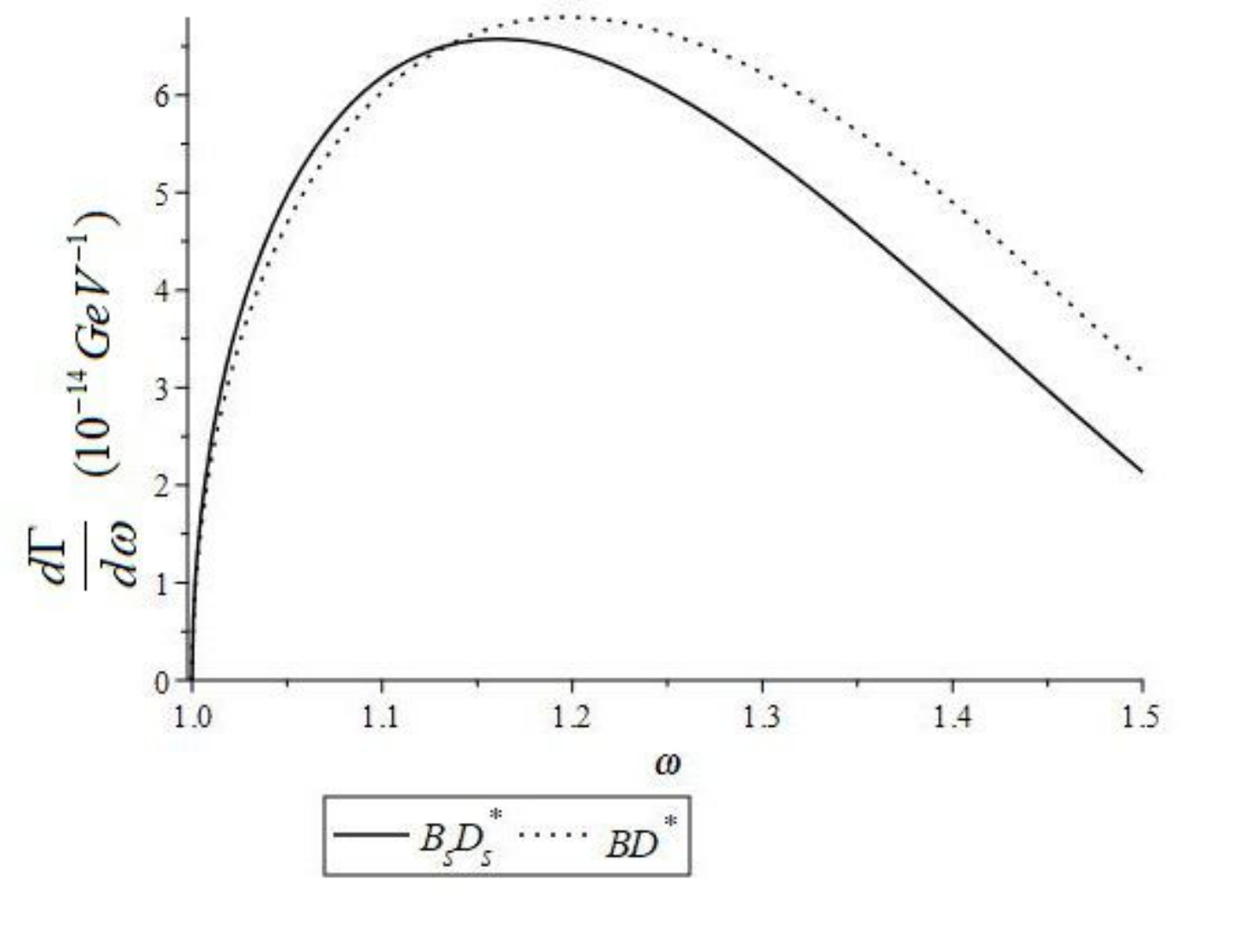}
\caption{Results of $\frac{{d\Gamma }}{{d\omega }}$ versus $\omega$.}
\label{fig:B to D_star}
\end{figure}

\begin{figure}
\centering
\includegraphics[width=0.48\linewidth]{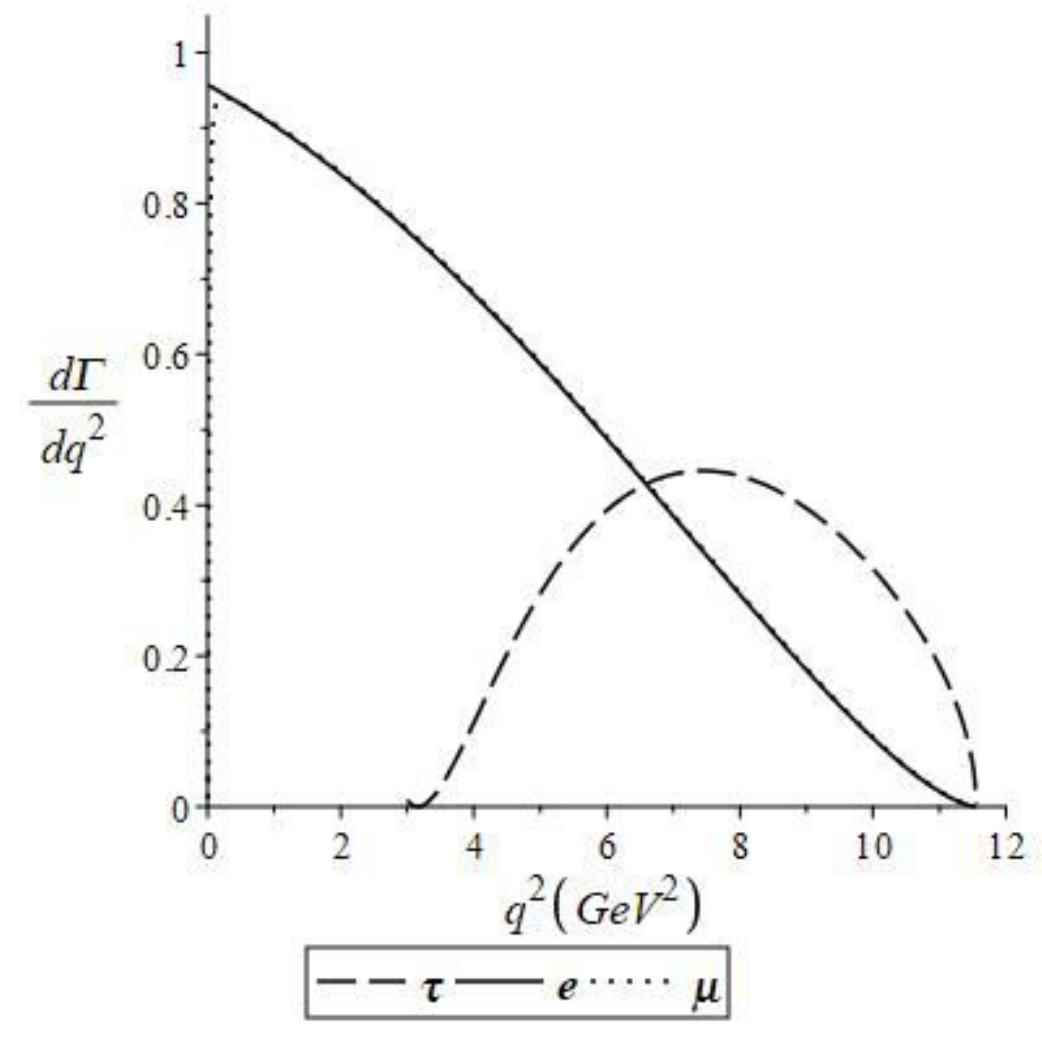}
\caption{Results of $\frac{{d\Gamma (B \to Dl{\nu _l})}}{{d{q^2}}} \times \frac{{{{10}^{12}}GeV}}{{|{V_{cb}}{|^2}}}$ versus transfer momentum.}
\label{fig:B to D leptons}
\end{figure}

For the case of pseudoscalar to vector $D$ and $D_s$ mesons, we can obtain the rates using the differential semileptonic decay widths  \cite{UKQCD:1995zee},
\begin{equation}
\label{eq:17}
    \begin{gathered}
  \frac{{d\Gamma (\bar B_{(s)}^{} \to D_{(s)}^*l\bar \nu )}}{{d\omega }} = \frac{{G_F^2}}{{48{\pi ^3}}}M_{D_{(s)}^*}^3{({M_{{B_{(s)}}}} - {M_{D_{(s)}^*}})^2}{[1 + {\beta ^{{A_1}}}(1)]^2} \times \sqrt {{\omega ^2} - 1} {(\omega  + 1)^2}|{V_{cb}}{|^2} \times  \hfill \\
  {\xi ^2}(\omega )\left[ {1 + \frac{{4\omega }}{{\omega  + 1}}\frac{{M_{{B_{(s)}}}^2 - 2\omega {M_{{B_{(s)}}}}{M_{D_{(s)}^*}} + M_{D_{(s)}^*}^2}}{{{{({M_{{B_{(s)}}}} - {M_{D_{(s)}^*}})}^2}}}} \right], \hfill \\ 
\end{gathered} 
\end{equation} 
with ${\beta ^{{A_1}}} =  - 0.01$. The dependence of $\omega $  for $\bar B_{(s)}^{} \to D_{(s)}^*l\bar \nu $  decay modes are shown in Fig. \ref{fig:B to D_star}. The solid and dotted lines, corresponding to ${\bar B_s} \to D_s^{* - }{l^ + }{\nu _l}$ and $\bar B \to {\bar D^{*0}}{l^ + }{\nu _l}$, respectively, show the differential decay rate $\frac{{d\Gamma }}{{d\omega }}$ dependence of the recoil variable in the kinematic region. The peaks of differential rate $\frac{{d\Gamma }}{{d\omega }}$ are placed at $\omega = 1.20$ GeV for $\bar B \to {\bar D^{*0}}{l^ + }{\nu _l}$ and $\omega = $ 1.16 GeV for ${\bar B_s} \to D_s^{* - }{l^ + }{\nu _l}$, respectively. In Fig. \ref{fig:B to D leptons}, $\frac{{d\Gamma (B \to Dl{\nu _l})}}{{d{q^2}}}$  is represented as a function of the momentum transfer squared for three lepton modes. As one can see from Fig. \ref{fig:B to D leptons}, the peak of differential decay width for $B \to D\mu {\nu _\mu }$ occurs before $q^2 = 2 GeV^2$. Our results of Fig. \ref{fig:B to D leptons} are in agreement with the results of Ref. \cite{Wang:2017jow}, where they showed the partial decay rates of $\frac{{d\Gamma (B \to Dl{\nu _l})}}{{d{q^2}}}$ for different ranges. Our results are also reasonable compared to the Belle measurements, where they obtained the highest values for partial decay rates in the range $q^2 = 0$ to 0.98 $GeV^2$ and 0.98 to 2.16 $GeV^2$ \cite{Belle:2015pkj}. For the electron and muon channels, the differential decay rate decrease with the enhancement of $q^2$, while for the tau channel,  $\frac{{d\Gamma (B \to D{\tau {\nu _\tau }})}}{{d{q^2}}}$ increase until $q^2 = 7.44$ $GeV^2$ and then diminish. In Ref. \cite{Wang:2017jow}, the peak was reported at 7.20 $GeV^2$ in the tau channel.

 We show our results for the branching ratios of $B$, $B_s$ and $B_c$ in Table \ref{tab:BR semi} using Eqs. (\ref{eq:13semi}, \ref{eq:16}, \ref{eq:17}), our obtained masses of mesons from the previous section as well as the mean life of $B$ mesons ${\tau _B} = 1.638ps,{\tau _{{B_s}}} = 1.527ps,{\tau _{{B_c}}} = 0.510ps$ \cite{PDG:2022}. The first column of Table \ref{tab:BR semi} stands for different decay modes, the second column shows our obtained values for the decay widths, and the third column is for the branching ratios. We compare our results with Refs. \cite{PDG:2022,Issadykov:2017wlb,CLEO:2002fch,Chen:2011ut,Colangelo:1999zn,Hernandez:2006gt,Wang:2012lrc} in the forth column. Based on these results, we compute the ratio of tau to muon and electron semileptonic decays as:
 \par
{$R(D) = \frac{{\mathcal{B}(B \to D\tau {\nu _\nu })}}{{\mathcal{B}(B \to De{\nu _e})}} = 0.448$, $R(D) = \frac{{\mathcal{B}(B \to D\tau {\nu _\nu })}}{{\mathcal{B}(B \to D\mu {\nu _\mu })}} = 0.448$, \\} 
 which are in agreement with the other results, $R(D) = \frac{{\mathcal{B}(B \to D\tau {\nu _\nu })}}{{\mathcal{B}(B \to Dl{\nu _l})}} = 0.440;$ $l = e,\mu $ \cite{BaBar:2012obs}, $R(D) = 0.300$ \cite{Na:2015kha}, 0.299 \cite{Bigi:2016mdz}, and $0.403 \pm 0.040 \pm 0.024$ \cite{HFLAV:2016hnz}.

\begin{center}
\begin{table}[ht]
\caption{Semileptonic decay widths and branching ratios of $B$, $B_s$ and $B_c$.}
\label{tab:BR semi}
\begin{tabularx}{\textwidth}{|X|X|X|X|} 
\hline    
 \textbf{Decay}  & \textbf{$\Gamma $ (GeV)} & \textbf{BR (Ours)} & \textbf{BR (Others)}  \\  \hline  
  $\bar B \to {\bar D^0}{\tau ^ + }{\nu _\tau }$ & {${\text{4}}{\text{.24}} \times {10^{ - 15}}$}  & {${\text{10}}{\text{.57}} \times {10^{ - 3}}$} & $(7.7 \pm 2.5) \times {10^{ - 3}}$ \cite{PDG:2022}\\  \hline 
 $\bar B \to {\bar D^0}{e^ + }{\nu _e}$ & {${\text{9}}{\text{.47}} \times {10^{ - 15}}$} & {2.36 \%}  & {$(2.30 \pm 0.09)$ \%}  \cite{PDG:2022}  \\  \hline
$\bar B \to {\bar D^0}{\mu ^ + }{\nu _\mu }$ & {${\text{9}}{\text{.47}} \times {10^{ - 15}}$} & {2.36 \%} & { $(2.30 \pm 0.09)$ \%}  \cite{PDG:2022} \\  \hline
$\bar B \to {\bar D^{*0}}{l^ + }{\nu _l}$ & {${\text{2}}{\text{.64}} \times {10^{ - 14}}$} & {6.10 \%} &  {$(4.97 \pm 0.12)$ \%  \cite{PDG:2022}, $(6.50 \pm 0.20 \pm 0.43)$ \% \cite{CLEO:2002fch}}  \\ \hline
${\bar B_s} \to D_s^{* - }{l^ + }{\nu _l}$ & {${\text{2}}{\text{.37}} \times {10^{ - 14}}$} & {5.50 \%} & { $(5.3 \pm 0.5)$ \% } \cite{PDG:2022}, (5.1-5.8) \% \cite{Chen:2011ut} \\ \hline
${\bar B_s} \to D_s^ - {\mu ^ + }{\nu _\mu }$ & {${\text{7}}{\text{.49}} \times {10^{ - 15}}$} & {1.74 \%} & {$(2.44 \pm 0.23)$ \%}  \cite{PDG:2022}, 1.4-1.7 \cite{Chen:2011ut} \\  \hline
${\bar B_s} \to D_s^ - {e^ + }{\nu _e}$ & {${\text{7}}{\text{.49}} \times {10^{ - 15}}$} & {1.74 \%} & 1.4-1.7 \cite{Chen:2011ut} \\  \hline
${\bar B_s} \to D_s^ - {\tau ^ + }{\nu _\tau }$ & {${\text{3}}{\text{.73}} \times {10^{ - 15}}$} & {${\text{8}}{\text{.65}} \times {10^{ - 3}}$} & ${\text{(4}}{\text{.7 - 5}}{\text{.5)}} \times {10^{ - 3}}$
\cite{Chen:2011ut} \\  \hline
${\bar B_c} \to {\eta _c}e{\bar \nu _e}$ &  {${\text{4}}{\text{.26}} \times {10^{ - 15}}$} & {0.33} $\%$ & 0.15 \cite{Colangelo:1999zn}, ${0.48^{ + 0.02}}$ \cite{Hernandez:2006gt} \\  \hline

${\bar B_c} \to {\eta _c}\mu {\bar \nu _\mu }$ & {${\text{4}}{\text{.26}} \times {10^{ - 15}}$} & {0.33 $\%$} & 0.15 \cite{Colangelo:1999zn}, ${0.48^{ + 0.02}}$ \cite{Hernandez:2006gt} \\  \hline
${\bar B_c} \to {\eta _c}\tau {\bar \nu _\tau }$ & {${\text{2}}{\text{.88}} \times {10^{ - 15}}$} & {0.22 $\%$} &  $0.17^{+0.01}$ \cite{Hernandez:2006gt}, 0.14 \cite{Wang:2012lrc}, 0.24 \cite{Issadykov:2017wlb} \\
\hline
\end{tabularx} 
\end{table}
\end{center}
\par

Further, we obtain $R({\eta _c})$  as $R({\eta _c}) = \frac{{\Gamma ({{\bar B}_c} \to {\eta _c}\tau {{\bar \nu }_\tau })}}{{\Gamma ({{\bar B}_c} \to {\eta _c}\mu {{\bar \nu }_\mu })}} = 0.677$, which is compatible with the one reported in Ref. \cite{Hernandez:2006gt}, $R({\eta _c}) = \frac{{\Gamma ({{\bar B}_c} \to {\eta _c}\tau {{\bar \nu }_\tau })}}{{\Gamma ({{\bar B}_c} \to {\eta _c}\mu {{\bar \nu }_\mu })}} = 0.{\text{452}}_{ - 0.030}^{ + 0.034}$. Our calculated result for the semileptonic decay mode $\Gamma ({\bar B_c} \to {\eta _c}\tau {\bar \nu _\tau })$  is {${\text{2}}{\text{.88}} \times {10^{ - 15}}$ GeV,} which is close to the result of $\Gamma ({\bar B_c} \to {\eta _c}\tau {\bar \nu _\tau }) = {2.46^{ + 0.07}} \times {10^{ - 15}}$ GeV \cite{Hernandez:2006gt}. 
The semileptonic decay widths of $B_c$ was obtained $\Gamma ({\bar B_c} \to {\eta _c}{e^ + }{\bar \nu _e}) = 2.1 \times {10^{ - 15}}$ GeV by Colangelo {\it{et al.}} \cite{Colangelo:1999zn} and reported $\Gamma ({\bar B_c} \to {\eta _c}l{\bar \nu _l}) = 5.9 \times {10^{ - 15}}$ GeV \cite{Ebert:2003cn} by Ebert {\it{et al}.}

\section{ Two-body nonleptonic decay widths of $B_c$  to charm mesons}
\label{sec4}
Since $B_c$ meson is the only heavy meson consisting of two heavy quarks with different open flavours ($b$ and $c$), the study of weak decays, as the only possible decay, of this meson is interesting and challenging. $B_c$ is stable against the strong and electromagnetic interactions. Both of its constituents are heavy and thus it can decay individually, which yield rich decay channels. Many weak decay modes of $B_c$ have been reported experimentally \cite{PDG:2022}. The tree level weak decay of $B_c$ can be justified in three modes: (i) the $b$ quark decays to $c$ and $u$ quarks while $c$ quark stands as a spectator; (ii) the $c$ quark decays to $s, d$ quarks while the $b$ quark' role is a spectator; (iii) the relatively suppressed
weak annihilation mode. The quark level transition for the transitions ${B_c} \to {\eta _c},J/\psi $ is $b \to c$ induced. Hence, we want to evaluate the nonleptonic two-body decay widths of $B_c$ in the factorization approximation \cite{Hernandez:2006gt}. In fact, the factorization approximation assumes that the nonleptonic decay amplitude reduces to the product of the form factors and corresponding decay constants \cite{Ebert:2003wc}. The invariant form factors for the semileptonic decay of $B_c$ can be defined by \cite{Dubnicka:2017job}
\begin{equation}
\begin{gathered}
  \mathcal{M}_{S = 0}^\mu  = {P^\mu }{f_ + }({q^2}) + {q^\mu }{f_ - }({q^2}), \hfill \\
  {
  \mathcal{M}_{S = 1}^\mu  = \frac{1}{{{M_{{B_c}}} + {M_2}}}\epsilon _\nu ^\dag\{  - {g^{\mu \nu }}Pq{A_0}({q^2}) + {P^\mu }{P^\nu }{A_ + }({q^2}) + {q^\mu }{P^\nu }{A_ - }({q^2}) }\hfill \\
  {
   + i{\varepsilon ^{\mu \nu \alpha \beta }}{P_\alpha }{q_\beta }V({q^2})\}} , \hfill \\ 
\end{gathered} 
\end{equation}
where $P = {p_{_{{B_c}}}} + {p_2}$ and $q = {p_{_{{B_c}}}} - {p_2}$. The relations of the form factors $f_ \pm (q^2)$, $A_0(q^2)$, $A_+(q^2)$, $A_-(q^2)$, and $V(q^2)$ will be defined later. All physical observables can be expressed by the helicity form factors $H_m$ which can be written in terms of invariant form factors for spin zero and spin one cases according to the following formulas \cite{Dubnicka:2017job},
\begin{equation}
\label{eq:helicity for spin 0}
    \begin{gathered}
  {H_t}({q^2}) = \frac{1}{{\sqrt {{q^2}} }}\{ (M_{{B_c}}^2 - M_2^2){f_ + }({q^2}) + {q^2}{f_ - }({q^2})\} , \hfill \\
  {H_ \pm } = 0, \hfill \\
  {H_0}({q^2}) = \frac{{2{M_{{B_c}}}|{{\vec p}_2}|}}{{\sqrt {{q^2}} }}{f_ + }({q^2}), \hfill \\ 
\end{gathered} 
\end{equation}
for ${0^ - } \to {0^ - }$ transitions and
\begin{equation}
    \begin{gathered}
  {H_t}({q^2}) = \frac{1}{{{M_{{B_c}}} + {M_2}}}\frac{{{M_{{B_c}}}|{{\vec p}_2}|}}{{{M_2}\sqrt {{q^2}} }}\{ (M_{{B_c}}^2 - M_2^2)({A_ + }({q^2}) - {A_0}({q^2})) + {q^2}{A_ - }({q^2})\} , \hfill \\
  {H_ \pm }({q^2}) = \frac{1}{{{M_{{B_c}}} + {M_2}}}\{  - (M_{{B_c}}^2 - M_2^2){A_0}({q^2}) \pm 2{M_{{B_c}}}|{{\vec p}_2}|V({q^2})\} , \hfill \\
  {H_0}({q^2}) = \frac{1}{{{M_{{B_c}}} + {M_2}}}\frac{1}{{2{M_2}\sqrt {{q^2}} }}\{  - (M_{{B_c}}^2 - M_2^2)(M_{{B_c}}^2 - M_2^2 - {q^2}){A_0} + 4M_1^2|{{\vec p}_2}{|^2}{A_ + }\},  \hfill \\ 
\end{gathered} 
\end{equation}
for ${0^ - } \to {1^ - }$ transitions. In these formulas the momentum of an outgoing meson with a mass $M_2$ is $|{\vec p_2}| = \sqrt {\lambda (M_{{B_c}}^2,M_2^2,{q^2})} /2{M_{{B_c}}}$. To obtain form factors, we use the IWF from previous section, ${f_ \pm (q^2) } = \xi (\omega )\frac{{{M_{{B_c}}} \pm {M_2}}}{{2\sqrt {{M_{{B_c}}}{M_2}} }}$ corresponding to spin zero transitions. Note that the heavy-to-heavy transition form factors between two pseudoscalar mesons are reduced to the IWF \cite{Isgur:1990yhj} and we have \cite{Quang:book}
\begin{equation}
\label{eq:23}
    V({q^2}) = {A_ + }({q^2}) = {A_ - }({q^2}) = \frac{{{{({M_{{B_c}}} + {M_2})}^2}}}{{4{M_{{B_c}}}{M_2}}}\xi (\omega ),
\end{equation}
regarding to vector mesons \cite{Quang:book}
\begin{equation}
\label{eq:24}
    V({q^2}) = \frac{{{A_0}({q^2})}}{{1 - \frac{{{q^2}}}{{{{({M_{{B_c}}} + {M_2})}^2}}}}}.
\end{equation}
Substituting Eq. (\ref{eq:23}) in Eq. (\ref{eq:24}), we arrive at
\begin{equation}
    {A_0}({q^2}) = \frac{{\left( {{{({M_{{B_c}}} + {M_2})}^2} - {q^2}} \right)}}{{4{M_{{B_c}}}{M_2}}}\xi (\omega ).
\end{equation}
The two-body nonleptonic decay widths of $B_c$ in terms of the helicity amplitudes can be expressed as \cite{Dubnicka:2017job},
\begin{equation}
\label{eq:twobodyB_c}
    \begin{gathered}
  \Gamma ({B_c} \to {\eta _c}{D_q}) = {N_W}{\{ {a_1}{f_{D_q^ - }}{M_{D_q^ - }}H_t^{{B_c} \to {\eta _c}}(M_{D_q^ - }^2) + {a_2}{f_{{\eta _c}}}{M_{{\eta _c}}}H_t^{{B_c} \to D_q^ - }(M_{{\eta _c}}^2)\} ^2}, \hfill \\
  \Gamma ({B_c} \to {\eta _c}D_q^{* - }) = {N_W}{\{ {a_1}{f_{D_q^{* - }}}{M_{D_q^{* - }}}H_0^{{B_c} \to {\eta _c}}(M_{D_q^{* - }}^2) - {a_2}{f_{{\eta _c}}}{M_{{\eta _c}}}H_t^{{B_c} \to D_q^{* - }}(M_{{\eta _c}}^2)\} ^2}, \hfill \\
  \Gamma ({B_c} \to J/\psi D_q^ - ) = {N_W}{\{  - {a_1}{f_{D_q^ - }}{M_{D_q^ - }}H_t^{{B_c} \to J/\psi }(M_{D_q^ - }^2) + {a_2}{f_{J/\psi }}{M_{J/\psi }}H_0^{{B_c} \to D_q^ - }(M_{J/\psi }^2)\} ^2}, \hfill \\
  \Gamma ({B_c} \to J/\psi D_q^{* - }) = {N_W}\sum\limits_{i = 0, \pm }^{} {{{\{ {a_1}{f_{D_q^{* - }}}{M_{D_q^{* - }}}H_i^{{B_c} \to J/\psi }(M_{D_q^{* - }}^2) + {a_2}{f_{J/\psi }}{M_{J/\psi }}H_i^{{B_c} \to D_q^{* - }}(M_{J/\psi }^2)\} }^2}},  \hfill \\ 
\end{gathered} 
\end{equation}
where
\begin{equation}
    {N_W} = \frac{{G_F^2}}{{16\pi }}\frac{{|{P_2}|}}{{M_{{B_c}}^2}}|{V_{cb}}V_{cq}^t{|^2},
\end{equation}
and ${f_D},{f_\eta },{M_D},{M_\eta }$ and ${M_{{B_c}}}$ are the decay constants and masses taken from section \ref{sec2} and $|{V_{cd}}| = 0.221,|{V_{cb}}| = 0.04,|{V_{cs}}| = 0.987$  are chosen from \cite{PDG:2022}. In Eq. (\ref{eq:twobodyB_c}), $H_t^{{B_c} \to {\eta _c}}(M_{D_q^ - }^2)$ means the helicity form factor ${H_t}({q^2})$, for ${B_c} \to {\eta _c}$ transition, which can be obtained by Eq. (\ref{eq:helicity for spin 0}) in the case of pseudoscalar to pseudoscalar transitions. $a_1$ and $a_2$ express as the color indices and are defined by the combination of the Wilson coefficients as ${a_1} = {C_2} + {C_4} + \zeta ({C_1} + {C_3})$  and ${a_2} = {C_1} + {C_3} + \zeta ({C_2} + {C_4})$  \cite{Dubnicka:2017job}, where $\zeta ({C_1} + {C_3}) = 1 / N_c$ is the color-suppressed parameter taken as zero. In our calculations, we take the values ${a_1} = 0.93,{a_2} =  - 0.27$ from Ref. \cite{Dubnicka:2017job}, in which the authors considered the numerical values of Wilson coefficients as $C_1 = -0.2632$, $C_2 = 1.0111$, $C_3 = -0.0055$, and $C_4 = -0.0806$ \cite{Descotes-Genon:2013vna}. In Ref. \cite{Ivanov:2006ni}, authors used ${a_1} = 1.14,{a_2} =  - 0.2$.
In Ref. \cite{Naimuddin:2012dy}, authors reported $\Gamma (B_c^ -  \to {\eta _c}{D^ - }) = {(0.32{a_1} + 0.24{a_2})^2} \times {10^{ - 15}}GeV$ \cite{Naimuddin:2012dy}. Our result is $\Gamma (B_c^ -  \to {\eta _c}{D^ - }) = {(0.{\text{49}}{a_1} + 0.{\text{29}}{a_2})^2} \times {10^{ - 15}}GeV$, which agrees with the one of Ref. \cite{Naimuddin:2012dy}. For the case of $B_c$ to $D_s$, $\Gamma (B_c^ -  \to {\eta _c}D_s^ - ) = {({\text{1}}{\text{.82}}{a_1} + {\text{1}}{\text{.6}}{a_2})^2} \times {10^{ - 15}}GeV$ was obtained in Ref. \cite{Naimuddin:2012dy}, which is close to what we have, {$\Gamma (B_c^ -  \to {\eta _c}D_s^ - ) = {({\text{2}}{\text{.50}}{a_1} + {\text{1}}{\text{.54}}{a_2})^2} \times {10^{ - 15}}GeV$}. For the cases $B_c$ to vector states, one can see disparities between different results in the theoretical literature. For example for the branching ratio of $B_c^ -  \to {\eta _c}D_s^{* - }$  mode, we find these results, 0.057 \cite{Colangelo:1999zn}, 0.149 \cite{Naimuddin:2012dy}, 0.26 \cite{Gouz:2002kk}. We evaluate $BR(B_c^ -  \to {\eta _c}D_s^{* - }) = 0.220$ \%. The same is occurred in the mode $B_c^ -  \to J/\psi D_s^{* - }$, in which we obtain 0.026 \%, and the results vary from 1.97 \% \cite{Gouz:2002kk} to 0.019 \cite{Colangelo:1999zn}. In fact, each of the two heavy quarks in $B_c$ meson can decay weakly and this affects the results of decay properties of theoretical approaches  \cite{ATLAS:2015jep}. Our results for decay widths and branching ratios of nonleptonic decays of $B_c$  in Table \ref{tab:BR nonleptonic Bc} are in a reasonable range in comparison with other theoretical results. From Table \ref{tab:BR nonleptonic Bc}, the dominant modes found in the category of $b \to c$ of $B_c$ meson decays are the ones $B_c^ -  \to {\eta _c}(D_s^ - ,D_s^{ - *})$ with predicted branching ratios of 0.283 \%  and 0.220 \%, respectively, which can be accessible experimentally at high-luminosity hadron colliders.

\begin{center}
\begin{table}[ht]
\caption{Decay widths and branching ratios of nonleptonic decays of $B_c$ (${a_1} = 0.93,{a_2} =  - 0.27$).}
\label{tab:BR nonleptonic Bc}
\begin{tabularx}{\textwidth}{|X|X|X|X|} 
\hline    
\textbf{Decay}  & \textbf{Decay width (GeV)} & \textbf{BR (in \%)} & \textbf{Others}  \\  \hline      
$B_c^ -  \to {\eta _c}{D^ - }$ & {${\text{1}}{\text{.40}} \times {10^{ - 16}}$} & {0.011}  & 0.019 \cite{Ivanov:2006ni}, 0.010 \cite{Anisimov:1998uk}, 0.015 \cite{Kiselev:2001zb}  \\  \hline
$B_c^ -  \to {\eta _c}D_s^ - $ & {${\text{3}}{\text{.65}} \times {10^{ - 15}}$} & {0.283} & 0.22 \cite{Dubnicka:2017job}, 0.21 \cite{Naimuddin:2012dy}\\  \hline 
 $B_c^ -  \to {\eta _c}{D^{* - }}$ & {${\text{1}}{\text{.13}} \times {10^{ - 16}}$} & {0.009}  & 0.0098 \cite{Dubnicka:2017job}, 0.003 \cite{Colangelo:1999zn}, 0.008 \cite{Naimuddin:2012dy}\\  \hline
 $B_c^ -  \to {\eta _c}D_s^{* - } $ & {${\text{2}}{\text{.84}} \times {10^{ - 15}}$} & {0.220} & 0.178 \cite{Naimuddin:2012dy}, 0.22 \cite{Dubnicka:2017job} \\  \hline
  $B_c^ -  \to J/\psi D_s^ - $ & {${\text{8}}{\text{.42}} \times {10^{ - 16}}$}  & {0.065}  & 0.041 \cite{Chang:1992pt} \\  \hline
$B_c^ -  \to J/\psi D_s^{* - }$ &  {${\text{3}}{\text{.30}} \times {10^{ - 16}}$} & {0.026} & 0.019 \cite{Colangelo:1999zn}, 0.41 \cite{Dubnicka:2017job} \\  \hline
$B_c^ -  \to J/\psi D_{}^ - $  & {${\text{4}}{\text{.08}} \times {10^{ - 17}}$} & {0.003} & 0.0035 \cite{Dubnicka:2017job}   \\  \hline

\end{tabularx} 
\end{table}
\end{center}

\begin{figure}
\centering
\includegraphics[width=0.48\linewidth]{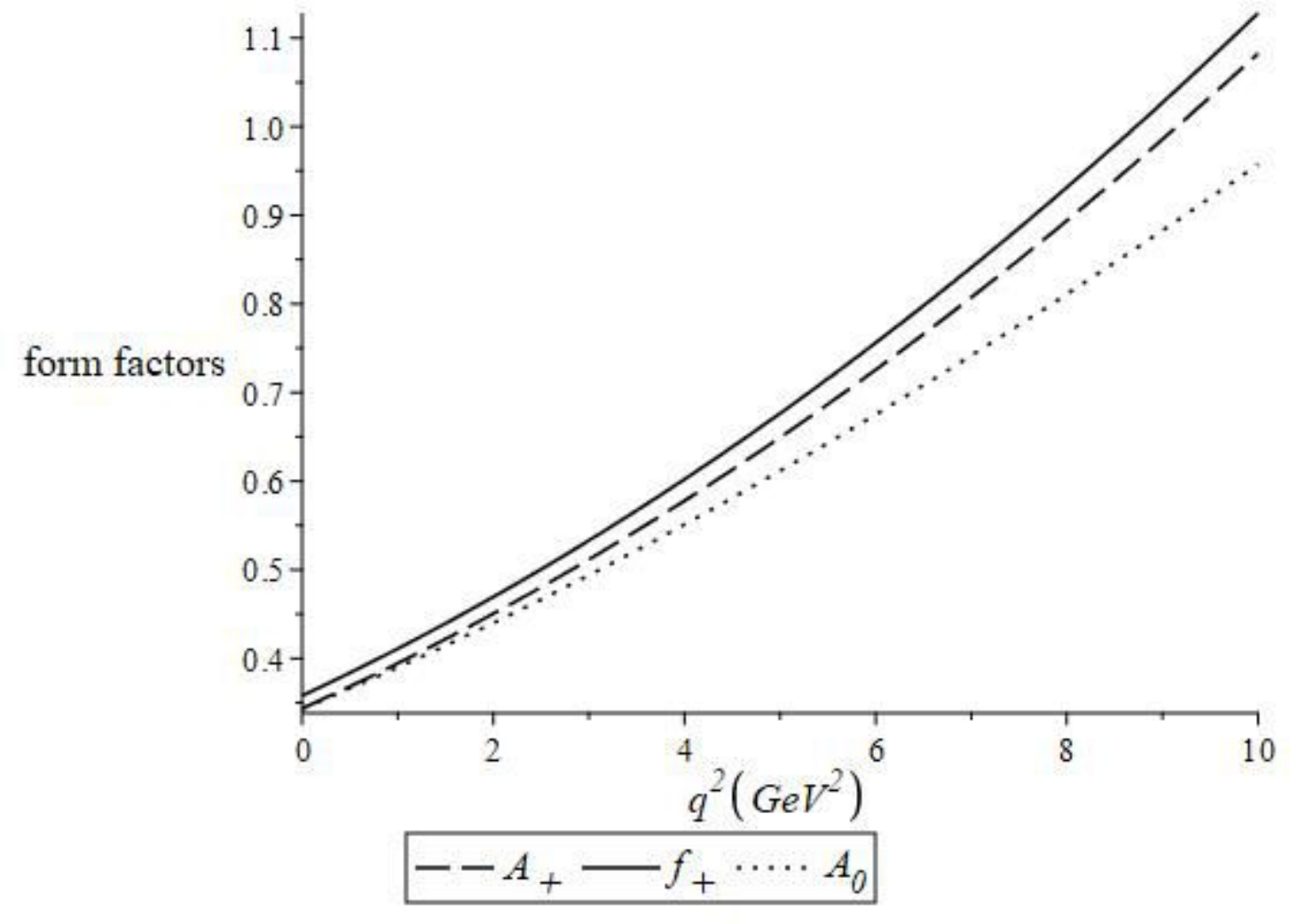}
\caption{Form factors for the decay $B_c^ -  \to J/\psi D_{}^ - $.}
\label{fig:Bc to J/psiD}
\end{figure}

\begin{figure}
\centering
\includegraphics[width=0.48\linewidth]{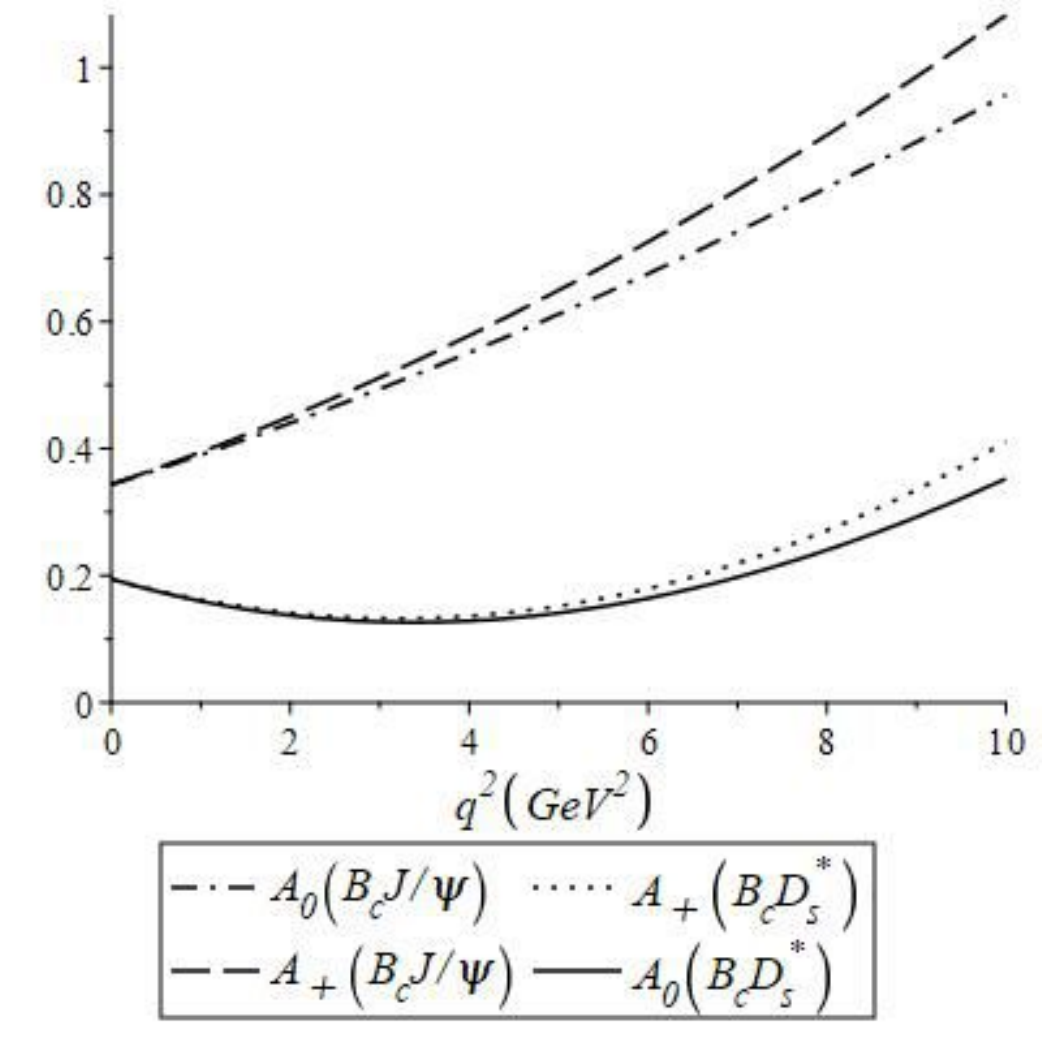}
\caption{Comparison of two form factors for the decay $B_c^ -  \to J/\psi D_s^{* - }$.}
\label{fig:Bc to J/psiD_star}
\end{figure}

\begin{figure}
\centering
\includegraphics[width=0.48\linewidth]{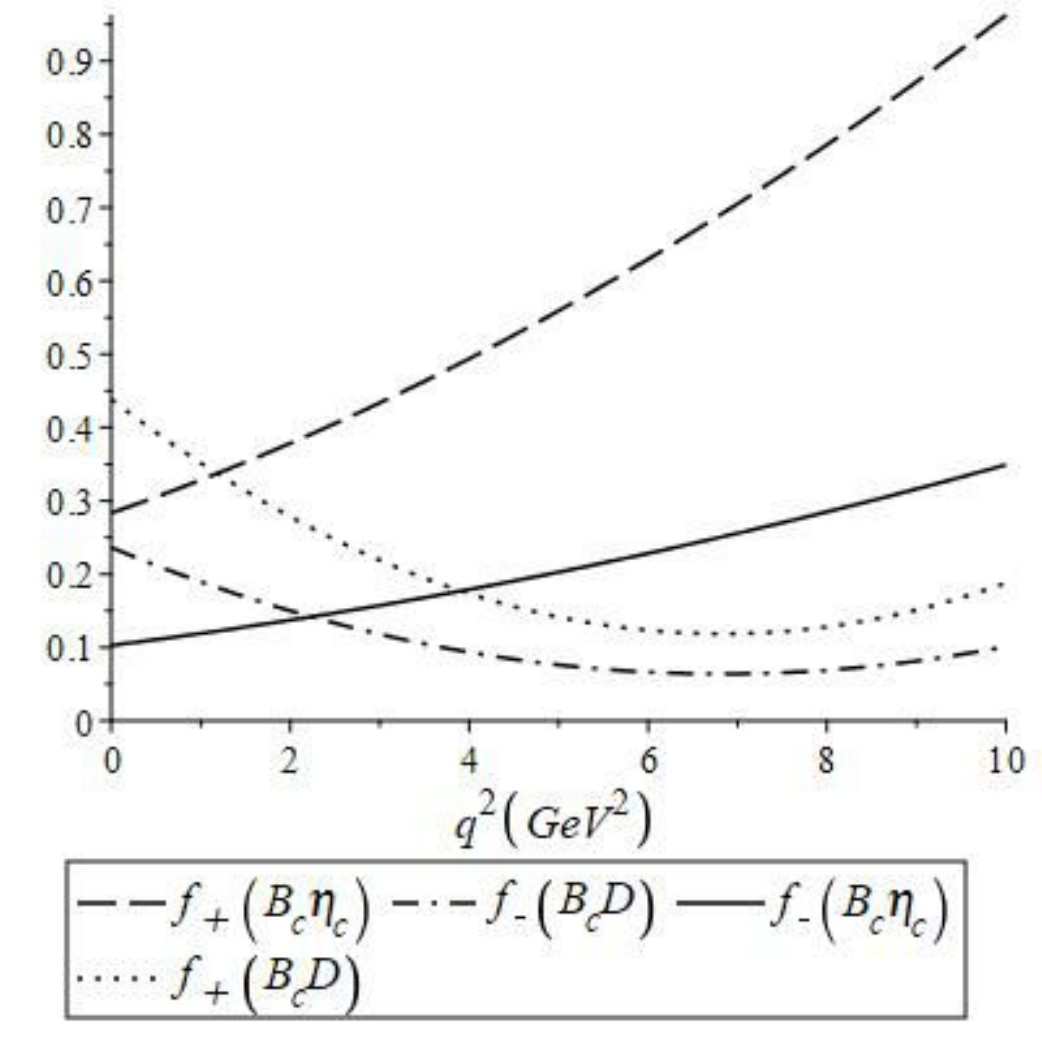}
\caption{Comparison of two form factors ${f_ \pm }$ for the ${B_c} \to \eta $ and ${B_c} \to D$ transitions .}
\label{fig:Bc to D and eta}
\end{figure}

In Fig. \ref{fig:Bc to J/psiD}, we plot form factors for the decay $B_c^ -  \to J/\psi D_{}^ - $, where the form factor ${f_ + }$ is related to ${B_c} \to D$ transition and ${A_0},{A_ + }$ stand for the form factors of ${B_c} \to J/\psi $. Fig. \ref{fig:Bc to J/psiD_star} represents the comparison of form factors for two decay modes in which the lower lines (solid and dotted) are related to ${B_c} \to D_s^*$  while the upper lines (dashed and dash-dotted) correspond to ${B_c} \to J/\psi $. We also show the comparison of two form factors ${f_ \pm }$ for the ${B_c} \to \eta $ and ${B_c} \to D$  transitions in Fig. \ref{fig:Bc to D and eta}.

\section{Conclusions}
\label{sec5}
It is still an important and interesting challenge to understand the dynamics of beauty and charm mesons. We present a non-relativistic quark model based on the Killingbeck potential to study different decay modes of $B$, $B_s$, and $B_c$. We employ a Gaussian wave function and obtained the mass spectrum in Tables \ref{tab:massb}, \ref{tab:massc} and the leptonic decay constants of bottom and charm mesons in Table \ref{tab:leptonic}. We evaluate the weak decay form factors for $B$ to $D$, $B_s$ to $D_s$, and the helicity amplitudes of $B_c$ to charm mesons using Isgur-Wise functions. The computed branching ratios for the semileptonic decay widths in Table \ref{tab:BR semi} and nonleptonic decay widths in Table \ref{tab:BR nonleptonic Bc} are in agreement with other available theoretical and experimental data, some predictions of which can be tested by the future experiments.


\begin{thebibliography}{10}

\bibitem{Neubert:1997uc}
{M.~Neubert and B.~Stech,
Adv. Ser. Direct. High Energy Phys. \textbf{15}, 294-344 (1998)
[arXiv:hep-ph/9705292 [hep-ph]].}


\bibitem{Ligeti:1993hw}
Z.~Ligeti, Y.~Nir and M.~Neubert,
Phys. Rev. D \textbf{49}, 1302-1309 (1994)
[arXiv:hep-ph/9305304 [hep-ph]].


\bibitem{Choi:2009ym}
H.~M.~Choi and C.~R.~Ji,
Phys. Rev. D \textbf{80}, 114003 (2009)
[arXiv:0909.5028 [hep-ph]].



\bibitem{Chen:2011ut}
X.~J.~Chen, H.~F.~Fu, C.~S.~Kim and G.~L.~Wang,
J. Phys. G \textbf{39}, 045002 (2012)
[arXiv:1106.3003 [hep-ph]].



\bibitem{Na:2015kha}
H.~Na \textit{et al.} [HPQCD],
Phys. Rev. D \textbf{92}, no.5, 054510 (2015)
[erratum: Phys. Rev. D \textbf{93}, no.11, 119906 (2016)]
[arXiv:1505.03925 [hep-lat]].



\bibitem{Dubnicka:2017job}
S.~Dubnicka, A.~Z.~Dubnickova, A.~Issadykov, M.~A.~Ivanov and A.~Liptaj,
Phys. Rev. D \textbf{96}, no.7, 076017 (2017)
[arXiv:1708.09607 [hep-ph]].



\bibitem{Zhou:2020ijj}
T.~Zhou, T.~Wang, Y.~Jiang, L.~Huo and G.~L.~Wang,
J. Phys. G \textbf{48}, no.5, 055006 (2021)
[arXiv:2006.05704 [hep-ph]].



\bibitem{ATLAS:2015jep}
G.~Aad \textit{et al.} [ATLAS],
Eur. Phys. J. C \textbf{76}, no.1, 4 (2016)
[arXiv:1507.07099 [hep-ex]].


\bibitem{Gershtein:1976aq}
S.~S.~Gershtein and M.~Y.~Khlopov,
Pisma Zh. Eksp. Teor. Fiz. \textbf{23}, 374-377 (1976)



\bibitem{Yang:2011ie}
M.~Z.~Yang,
Eur. Phys. J. C \textbf{72}, 1880 (2012)
[arXiv:1104.3819 [hep-ph]].



\bibitem{Mutuk:2018lki}
H.~Mutuk,
Adv. High Energy Phys. \textbf{2018}, 8095653 (2018)
[arXiv:1807.08511 [hep-ph]].



\bibitem{Gutierrez-Guerrero:2021fuj}
L.~X.~Gutierrez-Guerrero, J.~Alfaro and A.~Raya,
Int. J. Mod. Phys. A \textbf{36}, no.24, 2150171 (2021)
[arXiv:2108.12532 [hep-ph]].



\bibitem{Yang:2021crs}
Y.~Yang, Z.~Li, K.~Li, J.~Huang and J.~Sun,
Eur. Phys. J. C \textbf{81}, no.12, 1110 (2021)
[arXiv:2109.05650 [hep-ph]].



\bibitem{Kumar:2013dsa}
R.~Kumar and F.~Chand,
Commun. Theor. Phys. \textbf{59}, 528-532 (2013)


\bibitem{Hassanabadi:2014kka}
H.~Hassanabadi, S.~Rahmani and S.~Zarrinkamar,
Phys. Rev. D \textbf{89}, no.11, 114027 (2014)


\bibitem{Pang:2017dlw}
C.~Q.~Pang, J.~Z.~Wang, X.~Liu and T.~Matsuki,
Eur. Phys. J. C \textbf{77}, no.12, 861 (2017)
[arXiv:1705.03144 [hep-ph]].



\bibitem{Xiao:2020gry}
C.~W.~Xiao, S.~Rahmani and H.~Hassanabadi,
Eur. Phys. J. Plus \textbf{136}, no.10, 1083 (2021)
[arXiv:2007.03161 [hep-ph]].


\bibitem{Roberts:2007ni}
W.~Roberts and M.~Pervin,
Int. J. Mod. Phys. A \textbf{23}, 2817-2860 (2008)
[arXiv:0711.2492 [nucl-th]].

\bibitem{Bhaghyesh:2011zza}
Bhaghyesh, K.~B.~Vijaya Kumar and A.~P.~Monteiro,
J. Phys. G \textbf{38}, 085001 (2011)


\bibitem{VijayaKumar:2004hn}
K.~B.~Vijaya Kumar, B.~Hanumaiah and S.~Pepin,
Eur. Phys. J. A \textbf{19}, 247-250 (2004)


\bibitem{Pang:2019ttv}
C.~Q.~Pang,
Phys. Rev. D \textbf{99}, no.7, 074015 (2019)
[arXiv:1902.02206 [hep-ph]].


\bibitem{Lengyel:2000dk}
V.~Lengyel, Y.~Fekete, I.~Haysak and A.~Shpenik,
Eur. Phys. J. C \textbf{21}, 355-359 (2001)
[arXiv:hep-ph/0007084 [hep-ph]].



\bibitem{Radford:2009bs}
S.~F.~Radford, W.~W.~Repko and M.~J.~Saelim,
Phys. Rev. D \textbf{80}, 034012 (2009)
doi:10.1103/PhysRevD.80.034012
[arXiv:0903.0551 [hep-ph]].



\bibitem{PDG:2022}
R. L. Workman  {\it et al.}., (Particle Data Group),
Prog. Theor. Exp. Phys. \textbf{2022}, 083C01 (2022).


\bibitem{Braaten:1995ej}
E.~Braaten and S.~Fleming,
Phys. Rev. D \textbf{52}, 181-185 (1995)
[arXiv:hep-ph/9501296 [hep-ph]].


\bibitem{Belle:2006but}
K.~Ikado \textit{et al.} [Belle],
Phys. Rev. Lett. \textbf{97}, 251802 (2006)
[arXiv:hep-ex/0604018 [hep-ex]].



\bibitem{Sun:2019xyw}
H.~K.~Sun and M.~Z.~Yang,
Phys. Rev. D \textbf{99}, no.9, 093002 (2019)
[arXiv:1903.04295 [hep-ph]].


\bibitem{Pathak:2011km}
K.~K.~Pathak and D.~K.~Choudhury,
Chin. Phys. Lett. \textbf{28}, 101201 (2011)
[arXiv:1108.5315 [hep-ph]].


\bibitem{Eichten:2019gig}
E.~J.~Eichten and C.~Quigg,
Phys. Rev. D \textbf{99}, no.5, 054025 (2019)
[arXiv:1902.09735 [hep-ph]].


\bibitem{Ebert:2002pp}
D.~Ebert, R.~N.~Faustov and V.~O.~Galkin,
Phys. Rev. D \textbf{67}, 014027 (2003)
[arXiv:hep-ph/0210381 [hep-ph]].


\bibitem{Colquhoun:2015oha}
B.~Colquhoun \textit{et al.} [HPQCD],
Phys. Rev. D \textbf{91}, no.11, 114509 (2015)
[arXiv:1503.05762 [hep-lat]].


\bibitem{Wang:2005qx}
G.~L.~Wang,
Phys. Lett. B \textbf{633}, 492-496 (2006)
[arXiv:math-ph/0512009 [math-ph]].


\bibitem{Albertus:2005vd}
C.~Albertus, E.~Hernandez, J.~Nieves and J.~M.~Verde-Velasco,
Phys. Rev. D \textbf{71}, 113006 (2005)
[arXiv:hep-ph/0502219 [hep-ph]].




\bibitem{Ebert:2006hj}
D.~Ebert, R.~N.~Faustov and V.~O.~Galkin,
Phys. Lett. B \textbf{635}, 93-99 (2006)
[arXiv:hep-ph/0602110 [hep-ph]].


\bibitem{Chang:2018aut}
Q.~Chang, X.~N.~Li, X.~Q.~Li and F.~Su,
Chin. Phys. C \textbf{42}, no.7, 073102 (2018)
[arXiv:1805.00718 [hep-ph]].


\bibitem{Hassanabadi:2014isa}
H.~Hassanabadi, S.~Rahmani and S.~Zarrinkamar,
Eur. Phys. J. C \textbf{74}, no.10, 3104 (2014)
[arXiv:1407.3901 [hep-ph]].



\bibitem{Rahmani:2017vbg}
S.~Rahmani and H.~Hassanabadi,
Eur. Phys. J. A \textbf{53}, no.9, 187 (2017)



\bibitem{Atoui:2013zza}
M.~Atoui, V.~Mor\'enas, D.~Be\v{c}irevic and F.~Sanfilippo,
Eur. Phys. J. C \textbf{74}, no.5, 2861 (2014)
[arXiv:1310.5238 [hep-lat]].


\bibitem{Faller:2008tr}
S.~Faller, A.~Khodjamirian, C.~Klein and T.~Mannel,
Eur. Phys. J. C \textbf{60}, 603-615 (2009)
[arXiv:0809.0222 [hep-ph]].


\bibitem{LHCb:2020hpv}
R.~Aaij \textit{et al.} [LHCb],
JHEP \textbf{12}, 144 (2020)
[arXiv:2003.08453 [hep-ex]].


\bibitem{Wang:2017jow}
Y.~M.~Wang, Y.~B.~Wei, Y.~L.~Shen and C.~D.~L\"u,
JHEP \textbf{06}, 062 (2017)
[arXiv:1701.06810 [hep-ph]].



\bibitem{Isgur:1990yhj}
N.~Isgur and M.~B.~Wise,
Phys. Lett. B \textbf{237}, 527-530 (1990).


\bibitem{Xiao:2014ana}
Z.~J.~Xiao, Y.~Y.~Fan, W.~F.~Wang and S.~Cheng,
Chin. Sci. Bull. \textbf{59}, 3787-3800 (2014)
[arXiv:1401.0571 [hep-ph]].



\bibitem{Choi:2021mni}
H.~M.~Choi,
Phys. Rev. D \textbf{103}, no.7, 073004 (2021)
[arXiv:2102.02015 [hep-ph]].


\bibitem{UKQCD:1995zee}
K.~C.~Bowler \textit{et al.} [UKQCD],
Phys. Rev. D \textbf{52}, 5067-5094 (1995)
[arXiv:hep-ph/9504231 [hep-ph]].


\bibitem{Belle:2015pkj}
R.~Glattauer \textit{et al.} [Belle],
Phys. Rev. D \textbf{93}, no.3, 032006 (2016)
[arXiv:1510.03657 [hep-ex]].


\bibitem{Issadykov:2017wlb}
A.~Issadykov, M.~A.~Ivanov and G.~Nurbakova,
EPJ Web Conf. \textbf{158}, 03002 (2017)
[arXiv:1907.13210 [hep-ph]].


\bibitem{CLEO:2002fch}
N.~E.~Adam \textit{et al.} [CLEO],
Phys. Rev. D \textbf{67}, 032001 (2003)
[arXiv:hep-ex/0210040 [hep-ex]].



\bibitem{Colangelo:1999zn}
P.~Colangelo and F.~De Fazio,
Phys. Rev. D \textbf{61}, 034012 (2000)
[arXiv:hep-ph/9909423 [hep-ph]].



\bibitem{Hernandez:2006gt}
E.~Hernandez, J.~Nieves and J.~M.~Verde-Velasco,
Phys. Rev. D \textbf{74}, 074008 (2006)
[arXiv:hep-ph/0607150 [hep-ph]].



\bibitem{Wang:2012lrc}
W.~F.~Wang, Y.~Y.~Fan and Z.~J.~Xiao,
Chin. Phys. C \textbf{37}, 093102 (2013)
[arXiv:1212.5903 [hep-ph]].


\bibitem{BaBar:2012obs}
J.~P.~Lees \textit{et al.} [BaBar],
Phys. Rev. Lett. \textbf{109}, 101802 (2012)
[arXiv:1205.5442 [hep-ex]].
  


\bibitem{Bigi:2016mdz}
D.~Bigi and P.~Gambino,
Phys. Rev. D \textbf{94}, no.9, 094008 (2016)
[arXiv:1606.08030 [hep-ph]].



\bibitem{HFLAV:2016hnz}
Y.~Amhis \textit{et al.} [HFLAV],
Eur. Phys. J. C \textbf{77}, no.12, 895 (2017)
[arXiv:1612.07233 [hep-ex]].


\bibitem{Ebert:2003cn}
D.~Ebert, R.~N.~Faustov and V.~O.~Galkin,
Phys. Rev. D \textbf{68}, 094020 (2003)
[arXiv:hep-ph/0306306 [hep-ph]].


\bibitem{Ebert:2003wc}
D.~Ebert, R.~N.~Faustov and V.~O.~Galkin,
Eur. Phys. J. C \textbf{32}, 29-43 (2003)
[arXiv:hep-ph/0308149 [hep-ph]].

\bibitem{Quang:book}
H. K. Quang and X. Y. Pham, “Elementary Particles and Their Interaction,” Springer, 1998 

\bibitem{Descotes-Genon:2013vna}
S.~Descotes-Genon, T.~Hurth, J.~Matias and J.~Virto,
JHEP \textbf{05}, 137 (2013)
[arXiv:1303.5794 [hep-ph]].

\bibitem{Ivanov:2006ni}
M.~A.~Ivanov, J.~G.~Korner and P.~Santorelli,
Phys. Rev. D \textbf{73}, 054024 (2006)
[arXiv:hep-ph/0602050 [hep-ph]].








\bibitem{Naimuddin:2012dy}
S.~Naimuddin, S.~Kar, M.~Priyadarsini, N.~Barik and P.~C.~Dash,
Phys. Rev. D \textbf{86}, 094028 (2012)



\bibitem{Gouz:2002kk}
I.~P.~Gouz, V.~V.~Kiselev, A.~K.~Likhoded, V.~I.~Romanovsky and O.~P.~Yushchenko,
Phys. Atom. Nucl. \textbf{67}, 1559-1570 (2004)
[arXiv:hep-ph/0211432 [hep-ph]].



\bibitem{Anisimov:1998uk}
A.~Y.~Anisimov, I.~M.~Narodetsky, C.~Semay and B.~Silvestre-Brac,
Phys. Lett. B \textbf{452}, 129-136 (1999)
[arXiv:hep-ph/9812514 [hep-ph]].



\bibitem{Kiselev:2001zb}
V.~V.~Kiselev, O.~N.~Pakhomova and V.~A.~Saleev,
J. Phys. G \textbf{28}, 595-606 (2002)
[arXiv:hep-ph/0110180 [hep-ph]].



\bibitem{Chang:1992pt}
C.~H.~Chang and Y.~Q.~Chen,
Phys. Rev. D \textbf{49}, 3399-3411 (1994)




\end{thebibliography}
\end{document}